\documentclass[12pt, a4paper]{article}

\usepackage{amsmath}
\usepackage{amsfonts}
\usepackage{amssymb}
\usepackage{graphicx, rotating}
\usepackage{epstopdf}
\usepackage{epsfig}
\usepackage{latexsym}
\usepackage{graphicx}
\usepackage{color}
\usepackage{amsmath,bm,amssymb}
\usepackage{cite}
\usepackage{slashed}
\usepackage{hyperref}
\usepackage{arydshln}
\hypersetup{colorlinks, citecolor=bluscuro, linkcolor=black, urlcolor=bluscuro}
\definecolor{rossos}{cmyk}{0,1,1,0.55}
\definecolor{bluscuro}{rgb}{0.15, 0.2, .85}
\definecolor{bluchiaro}{cmyk}{1,.3,0.,0.1}

\usepackage[usenames,dvipsnames]{xcolor}

\usepackage{enumerate}

\newcommand{\eq}[1]{Eq.~(\ref{#1})}

\newcommand{\Dslash}{\!\not\!\!  D}
\newcommand{\be}{\begin{equation}}
\newcommand{\ee}{\end{equation}}
\newcommand{\bea}{\begin{eqnarray}}
\newcommand{\eea}{\end{eqnarray}}

\def\lra#1{\overset{\text{\scriptsize$\leftrightarrow$}}{#1}}
\def\bma#1{\mbox{\boldmath{$#1$}}}

\begin{document}

\begin{titlepage}
\today
\vspace{.3in}

\vspace{1cm}
\begin{center}
{\Large\bf  
Higgs windows to new physics
through $\bma{d=6}$ operators:\\[5mm]  
Constraints and one-loop  anomalous dimensions
}
\\
\bigskip\color{black}
\vspace{1cm}{
{\large  J.~Elias-Mir\'o$^{a,b}$, J.R.~Espinosa$^{b,c}$, E.~Masso$^{a,b}$, A.~Pomarol$^{a}$}
\vspace{0.3cm}
} \\[7mm]
   {\it {$^a$\, Dept.~de~F\'isica, Universitat Aut{\`o}noma de Barcelona, 08193~Bellaterra,~Barcelona}}\\
    {\it {$^b$\, IFAE, Universitat Aut{\`o}noma de Barcelona,
   08193~Bellaterra,~Barcelona}}\\
{\it $^c$ {ICREA, Instituci\'o Catalana de Recerca i Estudis Avan\c{c}ats, Barcelona, Spain}}
\end{center}
\bigskip

\vspace{.4cm}

\begin{abstract}

The leading contributions from  heavy new physics to Higgs processes can be captured in a model-independent way  by dimension-six operators  in an effective Lagrangian approach.
We present a complete analysis of how these contributions affect Higgs couplings. Under certain well-motivated assumptions, we find that 8 CP-even plus 3 CP-odd Wilson coefficients  parametrize the main impact in Higgs  physics,  as all other coefficients  are constrained by non-Higgs SM measurements. We calculate   the most relevant  anomalous dimensions  for   these Wilson coefficients,  which describe operator mixing from the heavy scale down to the electroweak scale.   
This allows us to find the leading-log corrections  to the predictions for the Higgs couplings  in specific models, such as the MSSM or composite Higgs, which we find to be significant in certain cases. 

\end{abstract}
\bigskip

\end{titlepage}

\section{Introduction}

The resonance observed at around 125 GeV at the LHC~\cite{Higgs}  has properties consistent with the Standard Model Higgs boson. 
More precise measurements of its couplings  will hopefully provide information on the origin of electroweak symmetry breaking (EWSB) in the Standard Model (SM). All natural  mechanisms proposed for EWSB introduce new physics at some scale $\Lambda$, not far from the TeV scale, that  generates  deviations in  the SM Higgs physics.
From this perspective, a convenient framework for a model-independent  analysis for these deviations is the effective Lagrangian  approach  that  consists 
in  enlarging  the SM Lagrangian by including  higher-dimensional operators built of SM fields   \cite{Buchmuller:1985jz}. These operators are the low-energy remnants of the heavy new physics integrated out at the scale $\Lambda$, 
which appears then as the scale suppressing these operators.

In this article we  perform a complete  study of the impact of the (dominant) dimension-six  operators
in the most  important Higgs couplings.
In particular, we calculate    the corrections  to single Higgs couplings, relevant
for the main Higgs decays and production mechanisms.
We will show that, for one family, there are  8 CP-even operators that can only affect Higgs physics  and no other SM processes (at tree-level).
This  corresponds to  the number of independent  dimension-six  operators that can be constructed with $|H|^2$, and implies that  
Higgs couplings to fermions, photons, gluons, and  $Z\gamma$ (for which large corrections are still possible) 
are characterised by independent Wilson coefficients.
The rest of  operators that could in principle affect Higgs physics at tree-level also 
enter in other SM processes and therefore can be constrained
by independent  (non-Higgs) experiments.
In our paper we will present  the main  experimental constraints on these operators, 
with a  full dedicated analysis  to be reported in Ref.~\cite{preparation}.
Our  article aims to  complete  part of the analysis of Higgs physics given in Refs.~\cite{Giudice:2007fh,Contino:2013kra}.

Out of the 8 CP-even operators that only affect Higgs physics, 5 of them are "tree-level" operators and  3 are "one-loop",  as we will explain. The 5 tree-level operators
affect directly the Higgs couplings to fermions, the kinetic term of the Higgs and the Higgs self-couplings. We calculate the anomalous dimensions of these 5 operators, which allow us to describe the renormalization group (RG) evolution of these Wilson coefficients from the heavy scale $\Lambda$, where they are generated, down to the electroweak-scale.\footnote{For the other 
 3 one-loop CP-even operators, as well as for the 3 one-loop CP-odd, the calculation of the main  anomalous dimensions has been given in \cite{GJMT,us}.}
We apply these results to find
 the leading-log corrections to the predictions for Higgs-couplings in several Beyond the Standard Model (BSM) scenarios: the Minimal Supersymmetric Standard Model (MSSM), universal theories (such as composite-Higgs models)
and models with a non-standard top.
We find that the corrections from this running can be  sizable for $\Lambda\sim$ few TeV,
and will become more relevant as we have better measurements of the Higgs couplings.
We  also calculate the anomalous dimensions of the operators contributing to the $S$ and $T$ parameters and to the $Zb\bar b$ couplings. The stringent experimental
constraints on these quantities can then be  translated into indirect bounds on Higgs operators.

The article is organized as follows. In Section~\ref{basis}  we review the classification of operators into two classes: tree-level and loop-induced operators. We also discuss the related issue of the election of the basis of effective dimension-six ($d=6$) operators. The modifications of the Higgs couplings as a function of the  Wilson coefficients are presented in Section~\ref{HiggsPhysics}.  Some of these Wilson coefficients can only be tested in Higgs physics, while the others are  constrained by LEP and Tevatron. In Section~\ref{sec:expconst} we show how to obtain such constraints.
In Section~\ref{sec:RGEs}  we calculate the renormalization group equations (RGE) of the Wilson coefficients most relevant for Higgs physics.
In Section~\ref{one-loop-examples} we illustrate the impact of such RG effects on particular scenarios of interest.
After this, we present our conclusions while some technical details are left for the Apendices.
Appendix~A deals with redundant operators and the field redefinitions that can be used to remove them from the Lagrangian.  Appendix~B presents the results for the one-loop anomalous dimensions of the  Wilson coefficients  before removing the redundant operators. Finally, Appendix~C
discusses the transformation properties of the $d=6$ operators under the custodial ${SU(2)_L}\otimes {SU(2)_R}$ symmetry, which can be helpful in understanding some of their properties ({\em e.g.}, concerning their mixing under RG evolution).

\section{Dimension-six  operator basis} 
\label{basis}

Let us consider  a  BSM  sector characterized by a new mass-scale  $\Lambda$
much larger than the electroweak scale $M_W$.
We will assume, among other requirements to be specified later, that this sector
preserves lepton and baryon number.
By integrating out this sector
and performing
  an expansion of SM fields and their derivatives $D_\mu$ over $\Lambda$,
we  obtain an effective  Lagrangian  made of local operators:
\begin{equation}
{\cal L}_{\rm eff}=\frac{\Lambda^4}{g^2_*}
{\cal L}\left(\frac{D_\mu}{\Lambda}\ ,\  \frac{g_H H}{\Lambda}\ ,\  \frac{g_{f_{L,R}}  f_{L,R}}{\Lambda^{3/2}}\ ,\ \frac{gF_{\mu\nu}}{\Lambda^2}\right)\simeq {\cal L}_4+{\cal L}_6+\cdots\, ,
\label{expansion}
\end{equation}
where ${\cal L}_d$ denotes  the  term in the expansion made of operators of dimension $d$.
By   $g_*$  we denote a generic coupling, while
 $g_H$ and $g_{f_{L,R}}$  are respectively accounting for the couplings of the Higgs-doublet  $H$  
 and SM fermions $f_{L,R}$ to the BSM sector, and  $g$ and $F_{\mu\nu}$ represent respectively the SM gauge couplings and field-strengths.\footnote{
With this we  are assuming that the SM gauge symmetry is also realized at energies above $\Lambda$
and therefore the couplings of the gauge bosons  to the BSM sector are the SM gauge couplings.
We can relax this  assumption by replacing $g$ by an arbitrary coupling.
}  The Lagrangian in~\eq{expansion} is based on dimensional analysis and the 
dependence on the couplings is easily obtained when the Planck constant $\hbar$  is put back in place.
All couplings introduced in   \eq{expansion}
can be  useful  as  bookkeeping parameters. 
In particular, a term in the Lagrangian that contains $n$ fields, will carry some coupling  
to the power $n-2$ (in this counting, $\lambda$, the Higgs quartic-coupling, is formally of order $g_*^2$).

The dominant effects of the BSM sector are  encoded in  ${\cal L}_6$, as
  ${\cal L}_4$  leads only to  an unphysical  redefinition of the SM couplings.
  There are different bases  used in the literature
for  the set of independent $d=6$ operators in  ${\cal L}_6$.
Although physics is independent of the choice of basis, it is clear that some bases
are better suited than others in order to extract the relevant information {\em e.g.} for Higgs physics.
A convenient feature to ask of a good basis is that it  captures  in  few operators  
 the impact of different   new-physics scenarios, at least for the most interesting cases.
    For example, in  universal theories,
 defined as those BSM scenarios whose corrections can be encoded  in operators made only of SM bosons,  the  bases used in Refs.~\cite{Giudice:2007fh,Hagiwara:1993ck} are appropriate since the physics effects can be captured by just  14 CP-even $d=6$ operators.
Therefore, 14 is the number of independent parameters  of the new physics effects and this number
must be the same in all  bases. However, the list of operators required to describe this same physics can contain many more than 14 operators in other bases, as for example in that of Ref.~\cite{Grzadkowski:2010es}. It follows that if we use such alternative bases  to study universal theories there will be correlations among operator coefficients, making the analysis more cumbersome.

Another important  consideration for the choice of basis 
is to  separate  operators  whose coefficients are  expected to have  different sizes
(again, at least in the main theories of interest).
For example, it is convenient to keep separated the
operators that can be induced at tree-level from integrating weakly-coupled states
from those that can only be generated at the one-loop level.
This helps in determining the  most relevant operators when dealing with a large class of BSM scenarios
such as supersymmetric, composite Higgs or  little Higgs models among others.
As shown in Ref.~\cite{us},  this criterium is also useful when considering one-loop operator mixing, since
 one finds that tree-level induced operators  often do not  contribute
to the RGE flow of one-loop induced ones, independently, of course, of the UV origin of the operators. 
In this particular sense, the basis of \cite{us} is better suited than that of \cite{Hagiwara:1993ck}. 
It is obvious that to meet
all the   criteria 
  given above we do not need to sacrifice generality (a main goal of this article), as long as one keeps a complete basis of operators, as we do.

The operators of our basis will be  broadly classified in  three classes \cite{Giudice:2007fh,us}.
The first two classes will consist of  operators that could in principle be generated at  tree-level when integrating out 
heavy  states with spin $\leq 1$  in renormalizable weakly-interacting theories.
 As we  show in Appendix~\ref{AppShifts}, these operators can be written as products of scalar,  fermion or vector currents of  dimension  less than 3.\footnote{This, together with the fact that field-redefinitions through equations of motion do not mix the two types of operators, makes the classification well defined and unambiguous.}
 Among these current-current operators we call  operators of the first class
those that involve extra powers of Higgs fields or SM fermions.
They will be  proportional to some power of the couplings $g_H$ or $g_{f_{L,R}}$, respectively. 
The importance of the operators of the first class is that they can be the most sizeable  ones
when  the theory is close to the  strong-coupling limit,    $g_H,g_{f_{L,R}}\sim 4\pi$.
Operators of the second class are instead those that involve extra (covariant) derivatives 
or gauge-field strengths and, according to \eq{expansion},  are generically suppressed by  
$1/\Lambda^2$ times a certain power of  gauge couplings.
 Finally,  in the  third class, we will have operators that cannot be generated
 from a tree-level exchange of heavy fields  
and  can only be induced,  in renormalizable weakly-coupled theories, at the one-loop level.
In this case, we expect    these operators to be   suppressed by   $g_*^2/(16\pi^2\Lambda^2)$.

We  then classify the $d=6$ operators as 
\begin{equation}
{\cal L}_{6} =\sum_{i_1}g^2_*\frac{c_{i_1}}{\Lambda^2}{\cal O}_{i_1}+\sum_{i_2}\frac{c_{i_2}}{\Lambda^2}{\cal O}_{i_2}+\sum_{i_3}\frac{\kappa_{i_3}}{\Lambda^2}{\cal O}_{i_3}\, ,
\label{6dim}
\end{equation}
where, for notational convenience, we introduce  the one-loop suppressed coefficients 
\be
\kappa_{i_3}\equiv \frac{g_*^2}{16\pi^2} c_{i_3}\, ,
\ee
for the third class  of operators.
 In weakly-coupled theories,   $c_i\sim  f_i(g/g_*,g_H/g_*,...)$, where $f_i(g/g_*,g_H/g_*,...)$ are functions that depend  on  ratios of couplings. 
We refer to the operators ${\cal O}_{i_1}$  and  ${\cal O}_{i_2}$ as "current-current" or "tree-level" operators,
while we call ${\cal O}_{i_3}$  "one-loop"  operators.\footnote{ For a classification of operators similar in spirit to ours, see \cite{AEW}.}

Although our basis follows a classification inspired by renormalizable weakly-coupled theories, it can also be useful
when dealing with   strongly-coupled BSM models.
For example,  if  the Higgs or SM fermions arise as  composite mesonic states of a 
strongly-interacting gauge theory with  no small parameter,
our basis  can  still give the right parametrization by  taking $g_*\sim 4\pi$.
  Also, strongly-coupled models that admit  a weakly-coupled holographic description 
 generate $d=6$ operators that follow the above classification.
In this case we have 
  $g_*\sim 4\pi/\sqrt{N}$ where $N$ plays the role of the number of colors of the strong sector.

Let us start defining our basis by considering first  operators made of SM bosons only \cite{Giudice:2007fh}.
In the first  class of operators, ${\cal O}_{i_1}$, we have
\begin{equation}
{\cal O}_H=\frac{1}{2}(\partial^\mu |H|^2)^2\ \  ,\ \ \
{\cal O}_T=
\frac{1}{2} (H^\dagger {\lra{D}_\mu} H)^2
 \ \ ,\ \ \
{\cal O}_r=|H|^2 |D_\mu H|^2\ \ ,\ \ \
{\cal O}_6=\lambda |H|^6\, .
\label{first6dim}
\end{equation}
Here we have defined $H^\dagger {\lra { D_\mu}} H\equiv H^\dagger D_\mu H - (D_\mu H)^\dagger H $, with
$D_\mu H = \partial_\mu H -i g\sigma^a W^a_\mu H/2 - i g' B_\mu H/2$ ($H$ is taken to have hypercharge $Y_H=1/2$).
For  ${\cal O}_6$, which involves six Higgs fields, an  extra  factor $g_*^2$ could be present. 
Nevertheless, we have substituted this by  $\lambda$,  the Higgs self-coupling
 defined as  $V=- m^2 |H|^2+\lambda |H|^4$. 
This is motivated by the fact that  
 the lightness of the Higgs suggests that there is a symmetry
protecting the Higgs  self-coupling to be of order $\lambda\sim m_h^2/(2v^2)\sim 0.13$.
  Examples are supersymmetry or  global symmetries
as in  composite Higgs models.

In the  second class of operators, ${\cal O}_{i_2}$,  we have
\begin{eqnarray}
&&{\cal O}_W=\frac{ig}{2}( H^\dagger  \sigma^a \lra {D^\mu} H )D^\nu  W_{\mu \nu}^a\ \ ,\ \ \
{\cal O}_B=\frac{ig'}{2}( H^\dagger  \lra {D^\mu} H )\partial^\nu  B_{\mu \nu}\, ,\nonumber\\
&&{\cal O}_{2W}=-\frac{1}{2}  ( D^\mu  W_{\mu \nu}^a)^2\ \ ,\ \ \
{\cal O}_{2B}=-\frac{1}{2}( \partial^\mu  B_{\mu \nu})^2\ \ ,\ \ \
{\cal O}_{2G}=-\frac{1}{2}  ( D^\mu  G_{\mu \nu}^A)^2\, .
\label{second6dim}
\end{eqnarray}
Since the last three operators involve two field strengths, we expect
  $c_{2W}\sim g^2/g^2_*$,
$c_{2B}\sim g^{\prime\, 2}/g^2_*$, and 
$c_{2G}\sim g^{2}_s/g^2_*$.

In the third class of operators, ${\cal O}_{i_3}$, we have the CP-even operators
\begin{eqnarray}
&&{\cal O}_{BB}={g}^{\prime 2} |H|^2 B_{\mu\nu}B^{\mu\nu} \ \ ,\ \ \
{\cal O}_{GG}=g_s^2 |H|^2 G_{\mu\nu}^A G^{A\mu\nu}\label{third6dim1}\, , \\
&&{\cal O}_{HW}=i g(D^\mu H)^\dagger\sigma^a(D^\nu H)W^a_{\mu\nu}\ \ ,\ \ \
{\cal O}_{HB}=i g'(D^\mu H)^\dagger(D^\nu H)B_{\mu\nu}\label{third6dim2}\, ,
\\
&&{\cal O}_{3W}= \frac{1}{3!} g\epsilon_{abc}W^{a\, \nu}_{\mu}W^{b}_{\nu\rho}W^{c\, \rho\mu}\ \ ,\ \ \
{\cal O}_{3G}= \frac{1}{3!} g_s f_{ABC}G^{A\, \nu}_{\mu}G^B_{\nu\rho}G^{C\, \rho\mu}\, ,
\label{third6dim3}
\end{eqnarray}
and the CP-odd operators
\begin{eqnarray}
\label{third6dimCP1}
&&{\cal O}_{B\widetilde B}={g}^{\prime 2} |H|^2 B_{\mu\nu}\widetilde B^{\mu\nu} \ \ ,\ \ \
{\cal O}_{G\widetilde G}=g_s^2 |H|^2 G_{\mu\nu}^A \widetilde G^{A\mu\nu}\, ,\\
\label{third6dimCP2}
&&{\cal O}_{H\widetilde W}=ig(D^\mu H)^\dagger\sigma^a(D^\nu H)\widetilde W^a_{\mu\nu}\ \ ,\ \ \
{\cal O}_{H\widetilde B}=ig'(D^\mu H)^\dagger(D^\nu H)\widetilde B_{\mu\nu}\, ,
\\
&&{\cal O}_{3\widetilde W}=\frac{1}{3!}g\epsilon_{abc}\widetilde W^{a\, \nu}_{\mu}W^{b}_{\nu\rho}W^{c\, \rho\mu}\ \ ,\ \ \
{\cal O}_{3\widetilde G}=\frac{1}{3!}g_s f_{ABC}\widetilde G^{A\, \nu}_{\mu}G^B_{\nu\rho}G^{C\, \rho\mu}\, ,
\label{third6dimCP3}
\end{eqnarray}
where $\widetilde F^{\mu\nu}=\epsilon^{\mu\nu\rho\sigma}F_{\rho\sigma}/2$.
There are two more CP-even operators involving two Higgs fields and gauge bosons,
${\cal O}_{WB}={g}^{\prime}g H^\dagger \sigma^a H W^a_{\mu\nu} B^{\mu\nu}$ 
 and
${\cal O}_{WW}={g}^2 |H|^2 W^a_{\mu\nu} W^{\mu\nu\, a}$ 
(and the equivalent CP-odd ones),
but  these can be eliminated using the identities~\footnote{For CP-odd operators the identities are $4{\cal O}_{H\widetilde B}+{\cal O}_{B\widetilde B}+{\cal O}_{W\widetilde B}=0$ and $4{\cal O}_{H\widetilde W}+{\cal O}_{W\widetilde W}+{\cal O}_{W\widetilde B}=0$.}
\begin{eqnarray}
&&
{\cal O}_B={\cal O}_{HB}+\frac{1}{4}{\cal O}_{BB}+\frac{1}{4}{\cal O}_{WB}\ ,\label{OpId1}\\
&&{\cal O}_W={\cal O}_{HW}+\frac{1}{4}{\cal O}_{WW}+\frac{1}{4}{\cal O}_{WB}\, .
\label{OpId2}
\end{eqnarray}
The operators ${\cal O}_{3 W}$ and ${\cal O}_{3 G}$ (and the corresponding CP-odd ones)
have three field-strengths  and then their corresponding coefficients  should scale as   $c_{3W}\sim g^2/g_*^2$
and $c_{3G} \sim g_s^2/g_*^2$ respectively.

Let us now examine $d=6$ operators  involving SM fermions, considering a single family to begin with.
Operators of the first class  involving the  up-type quark are
\bea
{\cal O}_{y_u}   &=&y_u |H|^2    \bar Q_L \widetilde{H}u_R\, ,\ \label{oy}
\nonumber\\
{\cal O}^u_{R} &=&
(i H^\dagger {\lra { D_\mu}} H)( \bar u_R\gamma^\mu u_R)\, ,\
\nonumber\\
{\cal O}^q_{L}&=&
(i H^\dagger {\lra { D_\mu}} H)( \bar Q_L\gamma^\mu Q_L)\, ,\ 
\nonumber\\
{\cal O}_{L}^{(3)\, q}&=&
(i H^\dagger \sigma^a {\lra { D_\mu}} H)( \bar Q_L\gamma^\mu\sigma^a Q_L)\, ,
\label{first6dimF}
\eea
where  $\widetilde{H}=i\sigma_2 H^*$,
and   in operators   $\propto \bar Q_Lu_R$ we include a Yukawa coupling  $y_u$   ($m_u=y_u v/\sqrt{2}$)
  as an order parameter of the chirality-flip. We also understand, here and in the following, that when
needed the Hermitian conjugate of a given operator is included in the analysis.
  In the first class we have, in addition, the four-fermion operators:
\bea
{\cal O}^q_{LL}   &=&  (\bar Q_L\gamma^\mu  Q_L)(\bar Q_L\gamma^\mu  Q_L)
\  ,\ \ \ \
{\cal O}_{LL}^{(8)\, q}   =  (\bar Q_L\gamma^\mu  T^A Q_L)(\bar Q_L\gamma^\mu  T^A Q_L)
\  ,   \nonumber \\
{\cal O}^u_{LR} &=&(\bar Q_L\gamma^\mu  Q_L)(\bar u_R\gamma^\mu  u_R)
\  ,\ \ \ \
{\cal O}_{LR}^{(8)\, u}=(\bar Q_L\gamma^\mu   T^AQ_L)(\bar u_R\gamma^\mu  T^A u_R)
\  ,   \nonumber \\
{\cal O}^u_{RR}&=& (\bar u_R\gamma^\mu   u_R)(\bar u_R\gamma^\mu  u_R)\, ,
\label{4f}
\eea
where $T^A$ are the $SU(3)_c$ generators. 
Other four-fermion operators are linear combinations
of the ones appearing in \eq{4f}; see for example \cite{Grzadkowski:2010es,AguilarSaavedra:2010zi}. 
Finally,  the
one-loop  (dipole) operators involving the up-type quark  are 
\bea
{\cal O}^u_{DB} &=&y_u\bar Q_L \sigma^{\mu \nu} u_R\, \widetilde H g'B_{\mu \nu}\ , \nonumber\\
{\cal O}^u_{DW}&=&y_u
 \bar Q_L \sigma^{\mu \nu} u_R\,  \sigma^a\widetilde HgW^a_{\mu \nu}\ , \nonumber\\
{\cal O}^u_{DG}&=&y_u
 \bar Q_L \sigma^{\mu \nu} T^A u_R\, \widetilde H g_sG^A_{\mu \nu}\, .
\label{third6dimF}
\eea

Similar operators  to those given above  can be  written for 
 the down-type quarks and leptons. For one family of fermions these  are given in Table~\ref{table:two}.
Among them,   there is a   new type of   operators,  involving    two different types of fermions, which, as we will see,  can have an  important impact on Higgs physics at the one-loop level. 
These are
\be
{\cal O}_{R}^{ud} =
y_u^\dagger  y_d(i \widetilde H^\dagger {\lra { D_\mu}} H)( \bar u_R\gamma^\mu d_R)
\, ,\label{Hud}
\ee
and 
\bea
{\cal O}_{y_uy_d}&=&y_u y_d (\bar Q_L^r   u_R)\epsilon_{rs}(\bar Q_L^s  d_R)
\ ,\qquad
{\cal O}_{y_uy_d}^{(8)}=y_u y_d (\bar  Q_L^r   T^A u_R)\epsilon_{rs} (\bar Q_L^s  T^A d_R)
\ , \nonumber\\ 
{\cal O}_{y_uy_e}&=&y_u y_e (\bar  Q_L^r   u_R)\epsilon_{rs} (\bar L_L^s e_R)
\ ,\qquad
{\cal O}^{\prime}_{y_uy_e}=y_u y_e (\bar  Q_L^{r\, \alpha}   e_R)\epsilon_{rs} (\bar L_L^s u^\alpha_R)
 \  ,\nonumber\\  
{\cal O}_{y_ey_d}&=&y_e y_d^\dagger (\bar L_L   e_R)(\bar  d_R Q_L)
\, ,
\label{4fud}
\eea
where $\epsilon=i\sigma_2$ and $\alpha$ labels color (only shown when contracted outside parentheses).
These operators are in principle of the first type.  Nevertheless  in  the four-fermion operators of \eq{4fud}  we have incorporated a product of  Yukawa couplings since they involve two chirality-flips, while in \eq{Hud} we  have also 
included Yukawa couplings as it is the case  in theories with a flavour symmetry, as discussed below. 
These operators  are  then only suppressed by $1/\Lambda^2$ as 
second-class operators.

There is some redundancy in the operators given above, as it is clear
that some of them can be eliminated by field redefinitions (see Appendix~\ref{AppShifts})
or using the equations of motion (EoM).
 For example,  the operator ${\cal O}_r$  can be eliminated 
by field redefinitions:
\begin{equation}
c_r {\cal O}_r\leftrightarrow  c_r\left[\frac{1}{2}\left({\cal O}_{y_u}+{\cal O}_{y_d}+{\cal O}_{y_e}+\text{h.c.}\right)-{\cal O}_{H}+2{\cal O}_{6}\right]\, .
\label{oy}
\end{equation}
Also,  we could eliminate all 5 operators
of \eq{second6dim}
by using the EoM for the gauge fields:
 \bea
 D^\nu  W_{\mu \nu}^a
&=&igH^\dagger  \frac{\sigma^a}{2} \lra {D_\mu} H +g\sum_f\bar f_L\frac{\sigma^a}{2} \gamma_\mu  f_L\, \nonumber ,\\ 
\partial^\nu  B_{\mu \nu}&=&ig' Y_H H^\dagger  \lra {D_\mu} H+g'\sum_f\left[ Y_{L}^f\bar f_L \gamma_\mu f_L+Y_{R}^f\bar f_R\gamma_\mu  f_R\right]\, \nonumber ,\\
 D^\nu  G_{\mu \nu}^A&=&g_s\sum_q \bar q \, T^A \gamma_\mu  q\, ,
\label{eom}
\eea
where $Y_{L,R}^f$ are the fermion hypercharges and $Y_H$ the Higgs hypercharge. 
In particular, we could trade ${\cal O}_B$ and ${\cal O}_W$  with other operators:
\bea
c_B{\cal O}_B&\leftrightarrow&  c_B\frac{g^{\prime\, 2}}{g^2_*}\left[-\frac{1}{2}{\cal O}_T+\frac{1}{2}\sum_f\left(Y_L^f {\cal O}^f_L+ Y_R^f{\cal O}^f_R\right)\right]\, , \nonumber\\
c_W{\cal O}_W&\leftrightarrow&  c_W\frac{g^2}{g^2_*}\left[-\frac{3}{2}{\cal O}_H+2{\cal O}_6+\frac{1}{2}\left({\cal O}_{y_u}+{\cal O}_{y_d}+{\cal O}_{y_e}+\text{h.c.}\right)+  \frac{1}{4}\sum_f{\cal O}_L^{(3)\, f}\right]\, ,
\label{tobg}
\eea
where, in the last expression, we have eliminated ${\cal O}_r$ using \eq{oy}.

\begin{table}
\begin{center}
\begin{tabular}{|c|}\hline
${\cal O}_H=\frac{1}{2}(\partial^\mu |H|^2)^2$\\
${\cal O}_T=\frac{1}{2}\left (H^\dagger {\lra{D}_\mu} H\right)^2$\\
${\cal O}_6=\lambda |H|^6$
  \\    \hline \hline 
${\cal O}_W=\frac{ig}{2}\left( H^\dagger  \sigma^a \lra {D^\mu} H \right )D^\nu  W_{\mu \nu}^a$\\
${\cal O}_B=\frac{ig'}{2}\left( H^\dagger  \lra {D^\mu} H \right )\partial^\nu  B_{\mu \nu}$\\
\hdashline
${\cal O}_{2W}=-\frac{1}{2}  ( D^\mu  W_{\mu \nu}^a)^2$\\
${\cal O}_{2B}=-\frac{1}{2}( \partial^\mu  B_{\mu \nu})^2$\\
${\cal O}_{2G}=-\frac{1}{2}  ( D^\mu  G_{\mu \nu}^A)^2$
 \\ \hline\hline
${\cal O}_{BB}={g}^{\prime 2} |H|^2 B_{\mu\nu}B^{\mu\nu}$\\
${\cal O}_{GG}=g_s^2 |H|^2 G_{\mu\nu}^A G^{A\mu\nu}$\\
\hdashline
${\cal O}_{HW}=i g(D^\mu H)^\dagger\sigma^a(D^\nu H)W^a_{\mu\nu}$\\
${\cal O}_{HB}=i g'(D^\mu H)^\dagger(D^\nu H)B_{\mu\nu}$\\
\hdashline
${\cal O}_{3W}= \frac{1}{3!} g\epsilon_{abc}W^{a\, \nu}_{\mu}W^{b}_{\nu\rho}W^{c\, \rho\mu}$\\
${\cal O}_{3G}= \frac{1}{3!} g_s f_{ABC}G^{A\, \nu}_{\mu}G^{B}_{\nu\rho}G^{C\, \rho\mu}$
 \\\hline
 \end{tabular}
\caption{\it 14 CP-even operators made of SM bosons. 
The operators are grouped in 3 different boxes corresponding to the 3 classes of operators defined in \eq{6dim}.
Dashed lines separate operators of different structure within a given class.
There are, in addition, the 6 CP-odd operators given in Eqs.~(\ref{third6dimCP1})-(\ref{third6dimCP3}).\label{table:one}}
\end{center}
\end{table}

For one family of fermions the set of operators that we use is collected in Tables~\ref{table:one} and \ref{table:two}.
 We keep  all operators of   Eqs.~(\ref{first6dim})-(\ref{third6dimCP3}), 
since they  are the relevant ones
for a well-motivated class of BSM scenarios such as universal theories,
with the exception of ${\cal O}_r$, that we eliminate of our basis using \eq{oy}. In Tables~\ref{table:one} and \ref{table:two} there are 58 operators;
adding the 6 bosonic CP-odd ones in Eqs.~(\ref{third6dimCP1})-(\ref{third6dimCP3}) leads to a total of 64 operators.
 We still have 5 redundant operators
that once eliminated leave a total of 59 independent operators, in agreement with \cite{Grzadkowski:2010es}. 
We leave free the choice of which 5 operators to eliminate: {\em e.g.}, the operators of \eq{second6dim} could be eliminated by using  \eq{eom} or, 
alternatively, we could trade 5 operators that contain fermions by the operators in \eq{second6dim}.
We will use later this freedom in different ways depending on  the physics process  studied.  
Other redundant operators are discussed in Appendix~\ref{AppShifts}.

{{\renewcommand{\arraystretch}{1.15} 
\begin{table}[htdp]
\begin{center}
\rotatebox{90}{
\begin{tabular}{|c|c|c|}
\hline 
${\cal O}_{y_u}   =y_u |H|^2    \bar Q_L \widetilde{H} u_R$     &
${\cal O}_{y_d}   =y_d |H|^2    \bar Q_L Hd_R$     &
${\cal O}_{y_e}   =y_e |H|^2    \bar L_L H e_R$
\\\hdashline
${\cal O}_{R}^u =
(i H^\dagger {\lra { D_\mu}} H)( \bar u_R\gamma^\mu u_R)$     &
${\cal O}_{R}^d =
(i H^\dagger {\lra { D_\mu}} H)( \bar d_R\gamma^\mu d_R)$       &
${\cal O}_{R}^e =
(i H^\dagger {\lra { D_\mu}} H)( \bar e_R\gamma^\mu e_R)$      
\\
${\cal O}_{L}^q=(i H^\dagger {\lra { D_\mu}} H)( \bar Q_L\gamma^\mu Q_L)$            & &
${\cal O}_{L}^l=(i H^\dagger {\lra { D_\mu}} H)( \bar L_L\gamma^\mu L_L)$  
\\
${\cal O}_{L}^{(3)\, q}=(i H^\dagger \sigma^a {\lra { D_\mu}} H)( \bar Q_L\gamma^\mu\sigma^a Q_L)$     & 
$$            &
${\cal O}_{L}^{(3)\, l}=(i H^\dagger \sigma^a {\lra { D_\mu}} H)( \bar L_L\gamma^\mu\sigma^a L_L)$    
 \\\hdashline
${\cal O}_{LR}^u =(\bar Q_L\gamma^\mu  Q_L)(\bar u_R\gamma_\mu  u_R)$      &
${\cal O}_{LR}^d =(\bar Q_L\gamma^\mu  Q_L)(\bar d_R\gamma_\mu  d_R)$         &
${\cal O}_{LR}^e =(\bar L_L\gamma^\mu  L_L)(\bar e_R\gamma_\mu  e_R)$\\
${\cal O}_{LR}^{(8)\, u}=(\bar Q_L\gamma^\mu   T^A Q_L)(\bar u_R\gamma_\mu  T^A u_R)$       &
${\cal O}_{LR}^{(8)\, d}=(\bar Q_L\gamma^\mu   T^A Q_L)(\bar d_R\gamma_\mu  T^A d_R)$       & 
\\
${\cal O}_{RR}^u= (\bar u_R\gamma^\mu   u_R)(\bar u_R\gamma_\mu  u_R)$        & 
${\cal O}_{RR}^d= (\bar d_R\gamma^\mu   d_R)(\bar d_R\gamma_\mu  d_R)$    &
${\cal O}_{RR}^e= (\bar e_R\gamma^\mu   e_R)(\bar e_R\gamma_\mu  e_R)$
\\
${\cal O}_{LL}^q  =  (\bar Q_L\gamma^\mu  Q_L)(\bar Q_L\gamma_\mu  Q_L)$         & 
&
${\cal O}_{LL}^l  =  (\bar L_L\gamma^\mu  L_L)(\bar L_L\gamma_\mu  L_L)$
\\
${\cal O}_{LL}^{(8)\, q}   =  (\bar Q_L\gamma^\mu  T^A Q_L)(\bar Q_L\gamma_\mu  T^A Q_L)$       &
&\\
\hdashline
${\cal O}_{LL}^{ql}  =  (\bar Q_L\gamma^\mu  Q_L)(\bar L_L\gamma_\mu  L_L)$&&\\
${\cal O}_{LL}^{(3)\, ql}  =  (\bar Q_L\gamma^\mu  \sigma^a Q_L)(\bar L_L\gamma_\mu \sigma^a L_L)$&&\\
${\cal O}_{LR}^{qe}  =  (\bar Q_L\gamma^\mu  Q_L)(\bar e_R\gamma_\mu  e_R)$&&\\
${\cal O}_{LR}^{lu}  =  (\bar L_L\gamma^\mu  L_L)(\bar u_R\gamma_\mu  u_R)$
&${\cal O}_{LR}^{ld}  =  (\bar L_L\gamma^\mu  L_L)(\bar d_R\gamma_\mu  d_R)$&\\
         ${\cal O}_{RR}^{ud}= (\bar u_R\gamma^\mu   u_R)(\bar d_R\gamma_\mu  d_R)$  &&  \\
 ${\cal O}_{RR}^{(8)\, ud}= (\bar u_R\gamma^\mu  T^A u_R)(\bar d_R\gamma_\mu  T^A d_R)$&&\\
${\cal O}_{RR}^{ue}  =  (\bar u_R\gamma^\mu  u_R)(\bar e_R\gamma_\mu  e_R)$    &    
${\cal O}_{RR}^{de}  =  (\bar d_R\gamma^\mu  d_R)(\bar e_R\gamma_\mu  e_R)$&\\
         \hline\hline
${\cal O}_{R}^{ud} =
y_u^\dagger  y_d(i \widetilde H^\dagger {\lra { D_\mu}} H)( \bar u_R\gamma^\mu d_R)$
 &&\\
 \hdashline
${\cal O}_{y_uy_d}=y_u y_d (\bar Q_L^r   u_R)\epsilon_{rs}(\bar Q_L^s  d_R)$&&\\
${\cal O}_{y_uy_d}^{(8)}=y_u y_d (\bar  Q_L^r   T^A u_R)\epsilon_{rs} (\bar Q_L^s  T^A d_R)$&&\\
${\cal O}_{y_uy_e}=y_u y_e (\bar  Q_L^r  u_R)\epsilon_{rs} (\bar L_L^s e_R)$&&\\
${\cal O}^{\prime}_{y_uy_e}=y_u y_e (\bar  Q_L^{r\, \alpha}   e_R)\epsilon_{rs} (\bar L_L^s u^\alpha_R)$&&\\
${\cal O}_{y_ey_d}=y_e y_d^\dagger (\bar L_L   e_R)(\bar  d_R Q_L)$&&
\\
\hline\hline
$ {\cal O}_{DB}^u =y_u\bar Q_L \sigma^{\mu \nu} u_R\, \widetilde H g'B_{\mu \nu}$  &
$ {\cal O}_{DB}^d =y_d\bar Q_L \sigma^{\mu \nu} d_R\,  H g'B_{\mu \nu}$  &
$ {\cal O}_{DB}^e =y_e\bar L_L \sigma^{\mu \nu} e_R\,  H g'B_{\mu \nu}$  
\\
${\cal O}_{DW}^u=y_u\bar Q_L \sigma^{\mu \nu} u_R\,  \sigma^a\widetilde HgW^a_{\mu \nu}$   &
${\cal O}_{DW}^d=y_d\bar Q_L \sigma^{\mu \nu} d_R\,  \sigma^a HgW^a_{\mu \nu}$   &
${\cal O}_{DW}^e=y_e\bar L_L \sigma^{\mu \nu} e_R\,  \sigma^a HgW^a_{\mu \nu}$  
\\
${\cal O}_{DG}^u=y_u
 \bar Q_L \sigma^{\mu \nu} T^A u_R\, \widetilde H g_sG^A_{\mu \nu}$ &
 ${\cal O}_{DG}^d=y_d
 \bar Q_L \sigma^{\mu \nu} T^A d_R\,  H g_sG^A_{\mu \nu}$ 
 &\\
\hline
  \end{tabular}
  }
\caption{\it 44 operators made of one-family of SM fermions. 
In the first column there are operators made of the up-type quark and other fermions; in the second column there are operators made only of the down-type quark and leptons; the third column lists operators made only  of leptons.
The operators are grouped in 3 different boxes corresponding to the 3 classes of operators defined in \eq{6dim}.
Dashed lines separate operators of different structure within a given class.}
\end{center}
\label{table:two}
\end{table}
}

Extending the  basis to 3 families increases  considerably the number of operators. 
We can reduce it by imposing  flavor symmetries, which are also 
needed to avoid tight constraints on flavor-violating processes.
For example,  we can require the BSM sector
to be invariant under  the flavor symmetry $U(3)_{Q_L}\otimes U(3)_{d_R}\otimes U(3)_{u_R} \otimes U(3)_{L_L}\otimes U(3)_{e_R}$,
under  which the corresponding 3 families transform as  triplets, and 
the Yukawas become $3\times 3$ matrices transforming as $y_d\in (\bf 3, \bar 3,0,0,0)$, 
$y_u\in (\bf 3,0, \bar 3,0,0)$ and $y_e\in (\bf 0,0,0,3, \bar 3)$ under the non-Abelian part of the flavor group. One can also assume that the Yukawas are the only source of CP violation.
This assumption goes under the name of Minimal Flavor Violation (MFV) \cite{MFV}.
In this case  the   list of  operators given in Table~2  can
be easily generalized to  include  3 families. For example, for operators involving two fermions, we have
\bea
( \bar L_L\gamma^\mu L_L)&\rightarrow& \left[\delta_{ij}+O(y_ey_e^\dagger/g^2_*)\right]( \bar L^i_L\gamma^\mu L^j_L)\ , \nonumber \\
y_e \bar L_L e_R&\rightarrow&  y^{ij}_e\left[1+O(y_e^\dagger y_e/g^2_*)\right]\bar L^i_L e^j_R\, ,
\eea
  ($i,j$ are family indices)  and similarly for other fermion species.
For  4-fermion operators, we have several possibilities to form singlets under the flavor group.
 For the leptons we  find four independent operators:
\bea
{\cal O}_{LL}^l   &=&  (\bar L^i_L\gamma^\mu  L^i_L)(\bar L^j_L\gamma_\mu  L^j_L)
\  ,\ \ \ \   \nonumber\\
{\cal O}_{LL}^{(3)\, l}   &=&  (\bar L^i_L\gamma^\mu  \sigma^a L^i_L)(\bar L^j_L\gamma_\mu  \sigma^a L^j_L)
\  ,   \nonumber \\
{\cal O}_{LR}^e &=&(\bar L^i_L\gamma^\mu  L^i_L)(\bar e^j_R\gamma_\mu  e^j_R)
\  ,   \nonumber \\
{\cal O}_{RR}^e &=& (\bar e^i_R\gamma^\mu   e^i_R)(\bar e^j_R\gamma_\mu  e^j_R)\, ,
\label{4leptons}
\eea
where we are neglecting terms of $O(y^2_e/g^2_*)$,
while the independent set of 4-quark operators can be found in the  Appendix of Ref.~\cite{Domenech:2012ai}. The MFV assumption that the Yukawas are the only source of CP violation implies that the Wilson coefficients are real. 
For the top quark, having a  Yukawa coupling  of order one,
departures from    flavor-universality      could be important.

It is useful, in order to understand what operators mix under the RGE, to 
derive the transformation of the coefficients (or equivalently, of the operators)
under the global custodial $SU(2)_L\otimes SU(2)_R$ symmetry and the parity $P_{LR}$
that interchanges $L\leftrightarrow R$.
A detailed analysis is given in  Appendix~\ref{appc}.
In Table~\ref{table:sym}
we present the quantum numbers of the coefficients of the tree-level operators involving the Higgs. 
{\renewcommand{\arraystretch}{1.3} 
\begin{table}
\centering
\small
\begin{tabular}{| c |  c | c |}
\hline 
\text{Spurion} & $SU(2)_L\otimes  SU(2)_R$ &  $P_{LR}$ \\  \hline \hline 
$y_f$& $\bm 2_R$  & \\ \hline
$g'$& ${\bm 3_R+\bm 1}$  & \\ \hline\hline
  $c_T$  & $  (\bm 3_R\otimes  \bm 3_R)_{s}$ &   \\ \hline
  $c_H,c_6$ & $\bm1$ & $+$ \\ \hline
 $c_B+c_W$ & $\bm1$ & $+$ \\ \hline
 $c_B-c_W$  & $\bm1$ & $-$ \\ \hline
     $c_{y_f} $ & $\bm1$ &  \\ \hline
        $ c_R^f $  & $\bm 3_R$ &  \\ \hline
         $ c_L^f$  & $\bm3_R$ &  \\ \hline
$c_L^{(3)\, f}$ & $\bm1$ & \\         \hline 
$c_{R}^{ud}$ & $\bm 1$ & \\         \hline
\end{tabular}
\caption{\it Quantum numbers under the custodial $SU(2)_L\otimes  SU(2)_R$ and left-right parity $P_{LR}$
of  the  SM couplings and coefficients of the   tree-level  operators  involving  Higgs fields.
We only show the $P_{LR}$-parities of the coefficients with a well-defined transformation, see Apendix~\ref{appc}.}
\label{table:sym}
\end{table}
}

\section{Higgs physics}
\label{HiggsPhysics}

Let us now describe the effects of the $d=6$ operators on Higgs physics. 
 We will only present the modifications of the Higgs couplings  important for single Higgs production and decay, working under the assumption of MFV, allowing however for CP-violating bosonic operators.
We split the relevant part of the Lagrangian in two parts,
\begin{equation}
\label{Lh}
{\cal L}_{h} = {\cal L}_{h}^{(0)} + {\cal L}_{h}^{(1)}\, .
\end{equation}
In ${\cal L}_{h}^{(0)}$ we keep the SM couplings and the 
effects of the current-current operators of   Tables~\ref{table:one} and \ref{table:two}, while ${\cal L}_{h}^{(1)}$ has the effects of the loop operators.
We can remove the momentum dependence from the Higgs couplings in ${\cal L}_{h}^{(0)}$ by using the EoM, 
so that we end up with Higgs  couplings at zero momentum.
After doing that, we have, in the canonical basis for the Higgs field $h$,
\bea
{\cal L}_{h}^{(0)} &=& g_{hff} \, h (\bar f_L f_R + {\rm h.c.}) \, + \, 
g_{hVV} \, h V^\mu V_\mu   \, + \,   g_{hZf_{L} f_{L}} \ h \, Z_\mu \bar f_{L} \gamma^\mu f_{L}  \nonumber  \\ &&
+ \, g_{hZf_{R} f_R} \ h \, Z_\mu \bar f_{R} \gamma^\mu f_{R}  + \,   g_{hWf_{L} f'_{L}} \ h \, W_\mu \bar f_{L} \gamma^\mu f'_{L}\, , 
\label{Lzero}
\eea
where a sum over fermions is understood and $V=W,Z$. 
The couplings read~\footnote{A coupling of $W_\mu^\pm$ to the right-handed current $\bar f_R\gamma^\mu f'_R$ is generated from the operator ${\cal O}^{ud}_R$ in \eq{Hud}, but we do not include it as  it is expected  to be suppressed by two Yukawa couplings (due to the MFV assumption) and hence to be small.} 
\bea
g_{hff}&=&g_{hff}^{\rm SM}\left[1-\left(\frac{c_H}{2}+c_{y_f}\right)\xi  + \frac{\delta G_F}{2 \, G_F} \right] \label{hff} \ ,  \nonumber  \\
g_{hWW}&=&g_{hWW}^{\rm SM}\left[1-\left(c_H-\frac{g^2}{g_*^2}c_W\right)\frac{\xi}{2}      + \frac{\delta G_F}{2 \, G_F} + 2 \, \frac{\delta M_W}{M_W} \right]\ ,  \nonumber  \\
g_{hZZ}&=&g_{hZZ}^{\rm SM}\left[1-\left(c_H -\frac{g^2}{g_*^2}c_Z      \right)\frac{\xi}{2}  - \widehat T   + \frac{\delta G_F}{2 \, G_F} \right]\ ,   \nonumber  \\
g_{hWf_Lf'_L}&=& \frac{1}{ 2 \sqrt{2} \, v}  \frac{g^3}{g^2_*} \, c_W \, \xi + \frac{2}{v}  \,   \delta g_W^{f_L } \  ,   \nonumber  \\
g_{hZf_Lf_L}&=& \frac{1}{2 v\cos\theta_W } \,
\frac{g^3}{g^2_*} \left( T_L^3c_Z-Q_f c_B\tan^2\theta_W \right) \xi
+ \frac{2}{v}  \, \delta g^{f_L}_{Z} \ ,  \nonumber  \\
g_{hZf_Rf_R}&=&- \, \frac{\tan^2\theta_W}{2 v\cos\theta_W } \, \frac{g^3}{g^2_*} \,
Q_f c_B \, \xi + \frac{2}{v}  \, \delta g^{f_R}_{Z}
\ .
\label{coups}
\eea
Here the SM couplings must be expressed  as  a function of  the input parameters $\alpha=e^2/(4\pi)$, the Fermi constant $G_F$ and the physical $m_h$, $M_Z$ and fermion masses. In these equations, $\theta_W$ is the weak mixing angle, $T^3_L=\pm1/2$ stands for the
weak isospin values of up and down components of $SU(2)_L$ fermion doublets, $Q_f$ is the fermion electric charge. We have defined
\be
\xi\equiv\frac{g^2_*v^2}{\Lambda^2}\ ,
\ee
with $v\simeq 246$  GeV, and
\be
c_Z = c_W + \tan^2 \theta_W \, c_B \ .
\ee
In the couplings of \eq{coups},  we have introduced
\bea
   \frac{\delta G_F}{G_F} &=& 2 \, \left[   c_{LL}^{ (3)\,  l}  \, - \, c_{L}^{(3)\,  l}  \, \right] \,\xi   \ ,   \label{GF} \\
    \frac{\delta M_W}{M_W} &=&      \, 
 \, \frac{1}{2(1-2\sin^2\theta_W)} \left[  \cos^2 \theta_W \, \widehat T - 2 \, \sin^2 \theta_W \, \widehat S 
 + \sin^2\theta_W \frac{\delta G_F}{G_F} \right]  \ ,  
\eea
and
\bea
\delta g_W^{f_L} &=& \frac{g}{\sqrt{2}} \, c_L^{(3)\, f}\, \xi \ , \nonumber \\
\delta g^{f_L}_{Z} &=& \frac{g}{2 \cos \theta_W} \,  (2T^3_L  c_L^{(3) \, f}-c^f_L) \, \xi , \nonumber \\
\delta g^{f_R}_{Z} &=& - \, \frac{g}{2 \cos \theta_W}  \,  c_R^f \, \xi \ . \label{delta_g}
\eea
Finally, we have made use of the precision electroweak parameters  \cite{Peskin:1990zt,Barbieri:2004qk}
\be
\widehat  S=(c_W+c_B)\frac{M^2_W}{\Lambda^2}\ ,\ \ \ 
\widehat  T=c_T\xi \  .
\label{st}
\ee
As we have stressed in the previous Section, not all the operators appearing in the Higgs couplings of \eq{coups} are independent. Once one has decided which are the  redundant operators which are not in the  basis, one should simply put equal to zero the corresponding operator coefficients.

The second term in the Lagrangian (\ref{Lh}) necessarily contains field derivatives. It reads
\bea
\label{loopy}
{\cal L}_{h}^{(1)} &=& g_{\partial h WW} \, (  W^{+\mu} W^-_{\mu \nu } \partial^\nu h + {\rm h.c.}) \, + \, 
g_{\partial h ZZ} \, Z^\mu Z_{\mu \nu } \partial^\nu h \, + \,  g'_{hZZ} \, h Z^{\mu \nu} Z_{ \mu \nu}
\nonumber \\
 &+&  g_{hAA} \, h A^{\mu \nu} A_{ \mu \nu} 
  \, + \,   g_{\partial h AZ} \,  Z^\mu A_{\mu \nu } \partial^\nu h
 \, + \,  g_{hAZ} \, h A^{\mu \nu} Z_{ \mu \nu}
\, + \,  g_{hGG} \, h G^{A \mu \nu} G^{A}_{ \mu \nu}\ ,
\eea
where we have defined $ V_{\mu\nu} = \partial_\mu V_\nu -  \partial_\nu V_\mu$, for $V=W^\pm,Z,A$. The couplings are given by
\bea
g_{\partial hWW}&=&  - \ \frac{g^2 v}{2 \Lambda^2} \,\kappa_{HW} \ ,  \nonumber  \\
g_{\partial hZZ}&=&  - \   \frac{g^2 v}{2 \Lambda^2} \, (\kappa_{HW} +   \kappa_{HB} \tan^2 \theta_W) \ ,  \nonumber  \\
    g_{hAA} &=&  \frac{e^2 v}{\Lambda^2} \, \kappa_{BB}  =      \frac{g'_{hZZ}}{\tan^2 \theta_W}  =  - \, \frac{g_{hAZ}}{2 \tan \theta_W}  \ ,  \nonumber  \\
g_{\partial hAZ}&=&   - \  \frac{g^2 v}{ 2 \Lambda^2} \,\tan \theta_W \,(\kappa_{HW} -  \kappa_{HB})  \ ,  \nonumber \\
g_{hGG}&=&  \frac{g_s^2 v }{\Lambda^2} \, \kappa_{GG} 
\, .
\label{coups3}
\eea
The contributions from the CP-violating bosonic operators can be easily obtained from \eq{loopy}
by replacing one of the  field strengths $F_{\mu\nu}$ in the operators  by $\widetilde F_{\mu\nu}$.
Only  the contributions from the dipole operators (third box of Table~\ref{table:two}) have been neglected
since they are assumed to be proportional to Yukawa couplings.

In the list of modified Higgs couplings (\ref{coups}), the tree-level operator $ {\cal O}_6$ does not play any role.
The simplest modified coupling containing this operator would be the triple Higgs vertex
\be
\delta {\cal L}_{h}^{(0)} =  
g_{hhh}^{\rm SM}
\left[1- \left(c_6 + \frac{3c_H}{2}\right)\xi    + \frac{\delta G_F}{2 \, G_F} \right]\,  h^3  \ ,
\ee
where $g_{hhh}^{\rm SM}$ is the SM value for the $h^3$ coupling.
Experimental access to this coupling is not yet possible. 

From the couplings in  Eqs.~(\ref{Lzero}) and (\ref{loopy}) 
it is easy to derive the  modifications of the
 main Higgs partial-widths  due to  $d=6$ operators
  \cite{Giudice:2007fh,Contino:2013kra,preparation}.\footnote{For loop-suppressed partial-widths, such as $h\to\gamma\gamma$, we remind the reader that $d=6$ operators can have an effect either directly or through modifications of the SM couplings that change the SM loop contribution to that particular decay \cite{Giudice:2007fh}.}
 The coefficients  $c^{(3)\, f}_L, c^f_L, c^f_R$ can also modify the  cross-section of $hf\bar f$ production, giving contributions
that grow with the energy.
A particularly interesting case is $pp\rightarrow qth$ ($q$ being a light quark) 
that is dominated by the subprocess $W_L b\rightarrow th$. At large energies 
this  grows with the energy as
\be
\left|{\cal A}(W_L b \rightarrow th)\right|^2\simeq  \left(\frac{4 g^2_* c_L^{(3)\, q_3}}{\Lambda^2}\right)^2  s(s+t)\, .
\ee
The  extraction of  new physics through this process has been studied in Ref.~\cite{Farina:2012xp}.


\section{Experimental constraints on the Wilson coefficients
\label{sec:expconst}}

As we saw in the previous Section, many $d=6$ operators can directly affect the Higgs couplings. 
Some of them   only  affect Higgs physics (at tree-level).  Their corresponding coefficients  are 
\be
\{
c_H,c_6,
c_{y_f},  \kappa_{BB},\kappa_{GG},   \hat\kappa_{WW}  , \kappa_{B\widetilde B},\kappa_{G\widetilde G},  \hat\kappa_{W\widetilde W} \}\, .
\label{relevantcis}
\ee
The reason for this is clear in the case of $c_H$ and $c_6$ as these operators contain exclusively Higgs fields;
and in the case of $ c_{y_f}, \kappa_{BB}$ and $\kappa_{GG} $ because, when the Higgs is substituted by its vacuum expectation value (VEV),
these operators simply lead to an innocuous renormalisation of SM parameters. 
The coefficient $\hat\kappa_{WW}$ corresponds to the  direction
 in  parameter space given by
 \footnote{In Ref.~\cite{Rujula:1991se}
these were called  blind directions, combinations of operators which a certain group of experiments cannot bound.
In the case of $\hat\kappa_{WW}$ that group is non-Higgs experiments.}
\be
\kappa_{HB}
=- \kappa_{HW}   = {4} \kappa_{BB} =c_W = - c_B\equiv 4\hat\kappa_{WW}\,  ,
\label{blind}
\ee
and the reason why this direction is only constrained by Higgs physics  is subtle in our basis.
The easiest way to see it is to go from our basis, that contains the subset
\be
\label{B1}
{\cal B}_1 = \{ {\cal O}_W, {\cal O}_{B}, {\cal O}_{HW}, {\cal O}_{HB} , {\cal O}_{BB} \}\ ,
\ee
to the basis containing the subset ${\cal B}_3$ defined in \cite{us}:
\be
{\cal B}_3 =\{ {\cal O}_W, {\cal O}_{B},   {\cal O}_{WW}, {\cal O}_{WB}, {\cal O}_{BB} \}\ .
\label{B2} 
\ee
One can go from
one to another using (\ref{OpId2}).  
Now, in the basis containing   $ {\cal O}_{WW}$ it is clear that its coefficient
cannot be bounded
by any non-Higgs SM processes, for exactly the same reasons as 
$\kappa_{BB}$.
We can now use  \eq{OpId2} to get the expression of ${\cal O}_{WW}$ in terms
of the operators in ${\cal B}_1$, 
\be
 {\cal O}_{WW} = 4( {\cal O}_W - {\cal O}_B ) - 4( {\cal O}_{HW} - {\cal O}_{HB} ) + {\cal O}_{BB}\, ,
\ee
which leads  to the direction given in \eq{blind}. Similarly, for the CP-odd operators,  $\hat\kappa_{W\widetilde W}$
corresponds to the direction:
\be
\kappa_{H\widetilde B}
=- \kappa_{H\widetilde W}   = {4} \kappa_{B\widetilde B}
\equiv 4  \hat\kappa_{W\widetilde W}
\, .
\label{blind2}
\ee
Although the coefficients $c_H,c_6$ and $c_{y_f}$ have no severe constraints from Higgs physics yet \cite{Falkowski:2013dza},
the coefficients $\kappa_{BB}$ and the difference $\kappa_{HW}-\kappa_{HB}$  are subject to strong constraints from $h\rightarrow\gamma\gamma$  and 
$h\rightarrow Z\gamma$ respectively (as these decays are one-loop suppressed in the SM). These give at 95\%CL\cite{Falkowski:2013dza}
\be
 - 0.0013 \lesssim  \frac{M_W^2}{\Lambda^2}  \kappa_{BB}  \lesssim 0.0018  \ ,\   \ \ \  
 - 0.016 \lesssim   \frac{M_W^2}{\Lambda^2}  ( \kappa_{HW}-\kappa_{HB} ) \lesssim 0.009  \, .
\label{hzg}
\ee
Notice that $\kappa_{HW}-\kappa_{HB}$ is odd under $P_{LR}$ [\eq{prlb}] and could be suppressed with respect to the sum
$\kappa_{HW}+\kappa_{HB}$ if the BSM sector  respects this parity.
Similarly, the coefficient  $\kappa_{GG}$ enters  in the production $GG\rightarrow h$ 
and  gets the bound\cite{Falkowski:2013dza}:
\be
  \frac{M_W^2}{\Lambda^2}  | \kappa_{GG} | \lesssim 0.004  \, .
\label{hgg}
\ee
The coefficients of the CP-odd operators enter quadratically in 
$\Gamma(h\rightarrow\gamma\gamma)$  and 
$\Gamma(h\rightarrow Z\gamma)$,  and therefore their effects are suppressed with  respect to CP-even ones.

Apart from the "Higgs-only" coefficients of \eq{relevantcis}, the rest of the coefficients of $d=6$ operators that enter in the Lagrangian of \eq{Lzero} and \eq{loopy}, relevant for single Higgs physics,
can in principle be constrained  by 
 (non-Higgs) SM processes.
 In the following we  present   the main experimental constraints  on these Wilson coefficients.
We  also discuss limits on other  Wilson coefficients that, although 
do not affect Higgs physics at tree-level, could do it at the one-loop level. 
The details of this study with a full dedicated quantitative analysis will be  presented in \cite{preparation}.
In what follows  we  assume MFV (unless explicitly stated)
and CP-invariance.

\subsection{Universal theories}

We  start considering  universal theories, leaving the generalization for later.
The new physics effects of these theories are captured by the  operators  listed in Table~\ref{table:one}.
Deviations in the $W^\pm$ and $Z^0$ propagators 
 can  be parametrized by four quantities,  $\widehat S,\widehat T,W$ and $Y$ 
 \cite{Barbieri:2004qk}.
The contributions from  $d=6$ operators to $\widehat S$ and $\widehat T$ 
have been written in (\ref{st}); the corresponding equations for $W$ and $Y$ read
\be
W=c_{2W}\frac{M^2_W}{\Lambda^2}\ ,\ \ \ 
Y=c_{2B}\frac{M^2_W}{\Lambda^2}\, .
\label{wy}
\ee
LEP1, LEP2 ($e^+e^-\rightarrow l^+ l^-$) and  Tevatron allow  to constrain independently each of these four  quantities,
all of them 
at the  per-mille level  \cite{Barbieri:2004qk}.\footnote{LHC data is also useful to constrain $W$, $Y$ and $c_{2G}$, which affect quark cross-sections at high energies 
\cite{Domenech:2012ai}.}
We saw in  (\ref{st}) that $\widehat S$ depends only on the combination $c_W+c_B$.
The gauge-boson part of the orthogonal combination, ${\cal O}_W-{\cal O}_B$, contains at least three gauge bosons 
\be
\big.({\cal O}_W -{\cal O}_B)  \big|_{\langle H\rangle}= O(V^3)\, ,
\label{WminusB}
\ee
and thus it is a blind direction for LEP1 experiments.
To constrain this direction, we have to consider
 the effect of $c_{W,B}$ on triple gauge-boson vertices, which can be cast in the form
\bea
\delta {\cal L}_{3V} &=& i g \cos \theta_W \left[ 
\delta g_1^Z \, Z^{\mu}  \left( W^{- \, \nu}  W^+_{\mu\nu} 
  - W^{+ \, \nu}   W^-_{\mu\nu} \right) \, 
 +   \delta \kappa_Z \, Z^{\mu\nu} W^-_\mu  W^+_\nu
 + \frac{\lambda_Z}{M^2_W} Z^{\mu\nu} W^{- \rho}_{\nu}  W^{+}_{\rho \mu}
 \right]
 \nonumber \\
&& + \, i g \sin \theta_W  \left[ \delta \kappa_\gamma \, A^{\mu\nu} W^-_\mu  W^+_\nu    
+ \frac{\lambda_\gamma}{M^2_W} A^{\mu\nu} W^{- \rho}_{\nu}  W^{+}_{\rho \mu}  \right] \, , 
\label{V3}
\eea
where again we have defined $ V_{\mu\nu} = \partial_\mu V_\nu -  \partial_\nu V_\mu$ for $V=W^\pm,Z,A$.
The contributions from $d=6$ operators to these couplings are  given by
\bea
 \delta g_1^Z &=&
 \frac{M^2_Z}{\Lambda^2}  (c_W + \kappa_{HW})   \ \nonumber  ,\\
  \delta \kappa_\gamma &=& 
    \frac{M^2_W}{\Lambda^2} (\kappa_{HW} + \kappa_{HB})\ ,\nonumber \\
 \delta \kappa_Z&=& \delta g_1^Z - \tan^2 \theta_W  \delta \kappa_\gamma
\ \nonumber  ,\\
  \lambda_Z &=&  \lambda_\gamma \ = \ \frac{M^2_W}{\Lambda^2} \kappa_{3W} \, ,
\label{3vert}
\eea
where we do not include a contribution from ${c}_{2W}$ since it is constrained to be small, as we have seen before.
The third relation, as well as the identity $\lambda_Z =  \lambda_\gamma$, are a consequence of limiting the analysis to  $d=6$ operators  \cite{Hagiwara:1993ck}. 
The best current limits on triple gauge-boson vertices still come from 
$e^+e^-\rightarrow W^+W^-$ at
LEP2 \cite{LEP2}, although LHC results are almost as good and will be better in the near future \cite{LHC_V3,LHC_V3_bis}. 
Leaving aside the contributions
from $\kappa_{3W}$,  that  we expect to be small in most theories in which the  SM gauge bosons are  elementary above $\Lambda$,
we  can use   the two-parameter fit  from LEP2 \cite{LEP2}
   which at 95\%CL reads
\bea
- 0.046 \leqslant & \delta g^Z_1 & \leqslant 0.050 \,  ,\nonumber \\
- 0.11 \leqslant & \delta \kappa_\gamma & \leqslant 0.084 \, .
\label{lep2}
\eea
These  are a factor $\sim 10$ weaker than  the constraints on the  coefficients $\widehat S$, $\widehat T$, $W$ and $Y$  from LEP1
 (for this reason we can neglect their contributions to 
 $e^+e^-\rightarrow W^+W^-$). As expected,
the two constraints in \eq{lep2} are orthogonal in parameter  space 
to the direction $\hat\kappa_{WW}$ of \eq{blind}, as can be seen  using \eq{3vert}. 
For this reason,
to obtain independent bounds on the 4 parameters $c_W$, $c_B$, $\kappa_{HB}$ and $\kappa_{HW}$,
 we need the constraint \eq{hzg} 
combined with  \eq{lep2}  and the bound on $\widehat S$. These bounds are at the percent level. 
In the  particular case of  $\kappa_i\ll c_i$, as expected in weakly-coupled theories, we  obtain the bound
\be
- 0.046 \lesssim \frac{M_Z^2}{\Lambda^2} c_W \lesssim 0.050\, .
\label{ccw}
 \ee
 
As we said, LHC tests of triple gauge-boson vertices are becoming comparable to those from LEP2,  and
it is foreseen that LHC will surpass LEP2 in these type of measurements \cite{LHC_V3,LHC_V3_bis}.
It follows that an important implication of our study is that  the LHC will have a direct impact on 
the improvement of the limits on  $c_W+\kappa_{HW}$, $\kappa_{HW}+\kappa_{HB}$ and
$\kappa_{3W}$. We will see in the next Subsection that this conclusion is also valid in non-universal theories.

\subsection{Non-universal theories}

Let us now  discuss  BSM   models without the universal assumption, considering then  all operators of the basis. 
We will follow a different strategy than in the previous Subsection.
Let us first look at electroweak leptonic physics
for which the experimental constraints are expected to be the strongest ones. Since we 
 assume MFV,  dipole operators (third box of Table~\ref{table:two})   
give corrections to SM processes   proportional to lepton masses and can then  be neglected.
We   use the redundancy in  our set of operators to
   eliminate, by using  \eq{tobg},  the 5 operators
${\cal O}_{2B,2W,2G}$, ${\cal O}^{(3)\, l}_{L}$ and  ${\cal O}^l_{L}$.
Taking $\alpha$, $M_Z$ and $G_F$ as input parameters, 
 the relevant operators for the leptonic data
are the 4 operators
 ${\cal O}_T$, ${\cal O}_W$, ${\cal O}_B$, ${\cal O}_R^e$ and 
the four-lepton operators of \eq{4leptons}.
LEP1 data and Tevatron  afford 4 well-measured experimental quantities: 
The charged-leptonic width $\Gamma(Z\rightarrow l^+l^-)$, 
the leptonic left-right asymmetry $A_{LR}^l$,  the  $Z$-width into neutrinos
 $\Gamma(Z\rightarrow \nu\bar \nu)=\Gamma_Z^{\rm total}-
\Gamma_Z^{\rm visible}$ and $M_W$.
These allow us to place bounds on  the 4 quantities $\{c_T,c_W+c_B, c^e_R,\delta G_F/G_F\}$
[where $\delta G_F/G_F$  is given in \eq{GF}]
at almost the same level as for universal theories.
 We again need the LEP2  constraint of \eq{lep2} from $e^+e^-\to W^+W^-$
to bound the difference $c_W-c_B$ [see \eq{WminusB}].
The only remaining operators are  four-lepton interactions 
but they can also be highly constrained from $e^+e^-\rightarrow l^+ l^-$ at LEP2.

Having these constraints in mind, we can now move to  the quark sector.
Higgs-fermion operators, as those in  \eq{first6dimF},   give  contributions to the gauge-boson couplings to quarks
that make them  depart from the leptonic ones by the amounts $\delta g_W^{q_L},  
\delta g^{q_L}_{Z}$ and $\delta g^{q_R}_{Z} $
given in (\ref{delta_g}).
Experiments put severe bounds on these deviations.
For example, we have limits at the per-mille level on deviations from  lepton-quark universality 
from $\beta$-decays and semileptonic $K$-decays \cite{Antonelli:2010yf}.
This implies that the coefficient $c^{(3)\, q}_L\,\xi$   can be constrained at this level.\footnote{The operator 
${\cal O}_{LL}^{(3)\, ql}=(\bar Q_L\gamma_\mu \sigma^a Q_L)( \bar L_L\gamma^\mu \sigma^a L_L)$ also
gives contributions to  $\beta$-decays and  $K$-decays,
but this can be independently  constrained by recent LHC data  \cite{Chatrchyan:2013lga}.}
For $c^q_L, c^u_R$ and $c^d_R$ the main constraints   come  from LEP1  measurements at the $Z$-pole.
These can put  bounds on   deviations of  the $Z$ couplings  to quarks, $\delta g_{Z}^{q_{L,R}}$, and 
on $c^q_{L}$ and $c^{u,d}_R$.  

As we saw, operators made of top quarks can depart from the MFV assumption
due to the large top Yukawa coupling.
If this is the case, we can still bound 
$(c_L^{q_3}+c_L^{(3)\, q_3})\xi$
from  the measurement of   the 
$Zb_L\bar b_L$  coupling 
at LEP1 which also gives a per-mille bound.
Interestingly,   a $P_{LR}$ symmetry can be imposed in the BSM sector
such that $c^{q_3}_L= \, - \, c_L^{(3)\, q_3}$ [see \eq{PLPRtobeused}],  allowing for  large deviations on $c_L^{q_3}-c_L^{(3)\, q_3}$. 
Recent LHC measurements of the $Wtb$ coupling  
\cite{LHCtop}
put some bounds on  $c_L^{(3)\, q_3}$  
but they are not very strong.  Also  $c_R^t$ has practically no bound due to the large uncertainty in the 
determination of the $Z t_R\bar t_R$ coupling \cite{LHCtop2}.    
Bounds on the Wilson coefficient $c_R^{tb}$, see \eq{Hud},  arise from $b\rightarrow s\gamma$ and read $-0.001\lesssim c_R^{tb}M^2_W/\Lambda^2\lesssim 0.006$  \cite{Vignaroli:2012si}.
These bounds will be improved in the future by the LHC.

Four-fermion operators involving quarks, as those in the first box of Table~\ref{table:two}, can also be constrained by
recent LHC data \cite{Domenech:2012ai}, while
the coefficients of the operators of the second  box of  Table~\ref{table:two}  have
no severe experimental constraints due to their  Yukawa suppression.
However, they  can affect Higgs physics  through operator mixing, as we will see in the next Section.
Finally,  bounds on dipole operators can be found, for example, in Ref.~\cite{Contino:2013kra}.

We conclude that, concerning the strength of experimental constraints,  we can distinguish  the following  sets of 
$d=6$ operators:
\begin{enumerate}[(i)]
\item First, we have those which can only affect  Higgs physics.
We have 8+3 operators of this type (CP-even plus CP-odd respectively) for one family, with real coefficients  given in \eq{relevantcis} \footnote{If we relax the MFV assumption that the $c_{y_f}$ are real, in addition to the 3 operators $\mathrm{Re}(c_{y_f})({\cal O}_{y_f}+{\cal O}_{y_f}^\dagger)$ we should also consider the 3 CP-odd operators $\mathrm{Im}(c_{y_f})({\cal O}_{y_f}-{\cal O}_{y_f}^\dagger)$.}.
As shown in Section~3, they  can  independently modify   the   Higgs decay-width to fermions, photons, gluons and  $Z\gamma$,
apart from a global rescaling of all Higgs amplitudes due to $c_H$.
 \item A second set of operators are 
those whose coefficients are severely restricted  by electroweak precision data, as explained above.
Eliminating, by  the EoM of \eq{tobg}, 
${\cal O}_{2B},{\cal O}_{2W},{\cal O}_{2G}$ and ${\cal O}_{L}^l$,${\cal O}_{L}^{(3)\, l}$,
these are  $c_W+c_B$ and $c_T$ that affect the $W/Z$ propagator, 
 and  $c_R^e,c_L^q, c_R^{u,d},c^{(3)\, q}_L$ that affect  $Vf\bar f$ vertices.
\item \label{iii}
In a third set, we have the operator coefficients that can affect the $ZWW/\gamma WW$  vertices and are, at present, constrained at the few per-cent level. These  are the  combinations  $\kappa_{HB}+\kappa_{HW}$ and  $c_{W}+\kappa_{HW}$
(and also $c_{3W}$ if we include $\lambda_Z$ in the analysis).
\end{enumerate}

We finally would like to mention that  our result is in  contradiction with Ref.~\cite{concha}   that obtained a smaller number of parameters to  characterize Higgs physics and triple gauge-boson vertices.  
The origin of this discrepancy is due to the following.
In our basis it is clear that  physics at LEP1  is   not sensitive to the blind direction
$c_W=-c_B$, since only the combination $c_W+c_B$ enters in the $\widehat S$ parameter.
This blind direction, however, becomes more complicated when one goes to other bases, such as  
that of Ref.~\cite{Grzadkowski:2010es}, in which 
 ${\cal O}_{W}$ and ${\cal O}_{B}$ are eliminated  [by using \eq{tobg}] in favor of  operators made of SM fermions.  In such bases
there is the  risk of   overestimating   the number of independent experimental constraints on the Wilson coefficients.

\section{Running effects from $\bma{\Lambda}$ to $\bma{M_W}$
\label{sec:RGEs}}
 
So far, we have implicitly assumed that the  Wilson coefficients were evaluated at the
electroweak scale, at which their effects can be eventually measured. However, particular UV completions
  predict the values of those coefficients at the scale $\Lambda$ where  the heavy BSM is integrated out.
The RG evolution from $\Lambda$ down to the electroweak scale, described by the corresponding anomalous dimensions, can be important in many cases.

Our main interest
is to calculate the anomalous dimensions of the  Wilson coefficients   that can have the largest impact on Higgs physics.
As we explained in the previous section, these are  the coefficients listed in \eq{relevantcis}.
 In Ref.~\cite{us} we already calculated the most relevant anomalous dimensions of   the $\kappa_i$  in \eq{relevantcis}. 
We showed  that   tree-level Wilson coefficients
do not enter, at the one-loop level, in the RGEs of  the
$\kappa_i$, a property that allowed us to complete the calculation of \cite{GJMT}
for the anomalous dimensions 
 relevant for $h\to \gamma\gamma,Z\gamma$. 
 In this section we extend  the analysis by calculating   the  anomalous dimensions for the 5 tree-level Wilson coefficients:
\be
\{
c_H,c_6,
c_{y_t},c_{y_b},c_{y_\tau}\}\, .
\label{cr}
\ee
We notice that even in the future, with   better measurements of the Higgs couplings,
and  then better bounds on \eq{cr}, 
 we still expect  \eq{cr} to give the main BSM contributions to Higgs physics, since
other Wilson coefficients, such as $c_W$, are expected to receive even stronger constraints from LHC (for a given $\Lambda$).

Generically, the anomalous dimensions are functions of other  Wilson coefficients:
\be
\gamma_{c_i}=\frac{d c_i}{d \log \mu} = \gamma_{c_i}(c_j)\ ,
\label{gammast}
\ee
where $\mu$ is the renormalization scale. 
In the RHS of \eq{gammast} we keep
 the $c_j$ coefficients that can potentially give   the most significant contributions to the RG running.
We keep the following $c_j$.
First, those of \eq{cr} as
they have no important experimental constraints and also are the most relevant  in BSM scenarios with $g_*$ large.
We also keep the Wilson coefficients   of operators involving the top quark, departing from the MFV assumption.
 These are 
 ${\cal O}^{q_3}_L$, ${\cal O}^t_R$, ${\cal O}^{(3)\, q_3}_L$ and   ${\cal O}^{tb}_R$, 
in addition to the  4-fermion  operators, ${\cal O}^{q_3}_{LL}$,
${\cal O}^{(8)\, q_3}_{LL}$, ${\cal O}^{t}_{LR}$,
${\cal O}^{(8)\, t}_{LR}$,  ${\cal O}_{y_ty_b}$, ${\cal O}^{(8)}_{y_ty_b}$, ${\cal O}_{y_ty_\tau}$ and  ${\cal O}'_{y_ty_\tau}$.
We have several motivations to keep them.
First, they have no large constraints from experiments.
Second, they  can induce large effects on the anomalous dimensions of \eq{cr}, since 
they are proportional to the  top Yukawa coupling.
Also their Wilson coefficients can be sizable in many  BSM models,
such as composite Higgs or supersymmetric theories, as we will discuss.
To summarize, we consider in the RHS of  \eq{gammast} the following Wilson coefficients:
\be
\{c_j\}=\{c_H,c_6,
c_{y_t},c_{y_b},c_{y_\tau},c_L,c_R,c_L^{(3)},c_R^{tb},c_{LL},
c_{LL}^{(8)},c_{LR},c_{LR}^{(8)},c_{y_ty_b},c^{(8)}_{y_ty_b},
c_{y_ty_\tau},c'_{y_ty_\tau}\}\ ,
\label{cjs}
\ee
where, from now on, we suppress the $q_3$ and $t$ superindices in the coefficients for simplicity.

We would like to mention that, even 
for those Wilson coefficients  that 
receive 
 experimental constraints, as those  discussed in the previous section,
 the fact that 
the constraints apply  to the ratios $c_j M_W^2/\Lambda^2$
means that  bounds  at the  percent-level
can allow for $c_j\sim O(1)$ if  $\Lambda \sim O({\rm TeV})$.
These coefficients could then also give potentially  
non-negligible effects in the $\gamma_{c_i}$.
An example of this is  $c_W$.  
Nevertheless, one can still expect that  the dominant effects 
will  be given by the coefficients in \eq{cjs} since, for a given $\Lambda$, they can  always be larger 
 than $c_W$. 

In addition, 
we will also extend our  calculation of   anomalous dimensions  
to other Wilson coefficients beyond \eq{cr}.
These correspond to    operators
constrained by the present experimental data,
and then their anomalous dimensions can be also useful to derive indirect   bounds on the coefficients of  \eq{cjs}.\footnote{
 Other  anomalous dimensions were calculated in \cite{Hagiwara:1993ck,Alam:1997nk}.}

 The anomalous dimensions presented below correspond to the basis of Tables~\ref{table:one} and \ref{table:two}, after using the five redundancies to eliminate the operators \{${\cal O}_L^l$, ${\cal O}_L^{(3)\, l}$, ${\cal O}_{RR}^e$, ${\cal O}_{LL}^l$, ${\cal O}_{RR}^{(8) \, d}$}\}. Nevertheless, removing or not these five operators and keeping the redundancy would not change our results (see Appendix \ref{AppRGEs} for more details).

\subsection{Anomalous dimensions of  operators relevant for Higgs physics\label{RGEs1}}

We  present here the anomalous dimensions for the Wilson coefficients in \eq{cr}, the ones expected to dominate deviations in Higgs physics, including the effects from the Wilson coefficients in \eq{cjs}. These are given by
\bea
16\pi^2\gamma_{c_H}&=&\left[
4N_c y_t^2 +24\lambda -\frac{3}{2}(3g^2+2{g'}^2)\right]c_H
+12N_c y_t^2 c_L^{(3)}\ ,
 \\[0.2cm]
16\pi^2\gamma_{\lambda c_6}&=&6\left[N_c y_t^2+18\lambda
-\frac{3}{4}(3g^2+{g'}^2)\right]\lambda c_6
+2(40\lambda-3g^2)\lambda c_H\nonumber\\
&&-16N_c\lambda y_t^2c_L^{(3)}+ 8N_cy_t^2(\lambda-y_t^2)c_{y_t} \ ,
\\[0.2cm]
16\pi^2\gamma_{c_{y_t}}&=& 
 \left[(4N_c+9) y_t^2+24\lambda-\frac{3}{2}(3g^2+{g'}^2)
\right]c_{y_t}+\left(3y_t^2+2\lambda-\frac{3}{2}g^2\right)c_H \nonumber\\
&&+
(2y_t^2+4\lambda-3g^2-{g'}^2)c_R-
2(y_t^2+2\lambda+2{g'}^2)c_L
\nonumber \\
&&+
4(-N_c y_t^2+3\lambda+{g'}^2)c_L^{(3)}+
8(y_t^2-\lambda)\left[c_{LR}+C_Fc_{LR}^{(8)}\right]\  ,
\label{ct}
\\[0.2cm]
16\pi^2\gamma_{c_{y_b}}&=& 
 \left[ 2(N_c+1)y_t^2+24\lambda-\frac{3}{2}(3g^2+{g'}^2)
\right]c_{y_b}+\left( 2\lambda-\frac{3}{2}g^2\right)c_H  + (2 N_c -1 ) y_t^2c_{y_t}\nonumber\\
&&+
2(2\lambda+{g'}^2)c_L +2\left[
(3-2N_c) y_t^2+6\lambda+{g'}^2\right]c_L^{(3)}    - 4 \frac{y_t^2}{g_*^2} \left(y_t^2+2 \lambda-\frac{3}{2}g^2
\right) c^{tb}_{R} \nonumber\\
 &&+2\frac{y_t^2}{g_*^2} ( \lambda- y_t^2)\left[\left(2N_c + 1\right)  c_{y_t y_b}+
C_F c^{(8)}_{y_ty_b}\right] \  , 
 \label{cb}
\eea
\bea
16\pi^2\gamma_{c_{y_\tau}}&=& 
 \left[2 N_c y_t^2+24\lambda-\frac{3}{2}(3g^2+{g'}^2)
\right]c_{y_\tau}+\left(2 \lambda-\frac{3}{2}g^2\right)c_H+
2 N_c y_t^2[c_{y_t}-2 c_L^{(3)}] \nonumber\\
&&-  2\frac{y_t^2}{g_*^2} N_c ( \lambda- y_t^2) \left( 2 c_{y_t y_\tau}+c^{\prime}_{y_t y_\tau}  \right) \ ,
\label{ctau}
\eea
where $N_c=3$ is the number of colors and $C_F=(N_c^2-1)/(2N_c)$. 
 Parametrically one has $\gamma_{c_i} \sim g_j^2 c_j/16\pi^2$ and we only keep $g_j^2 =\{y_t^2, g_s^2, g^2, {g'}^2,\lambda\}$, dropping $g_j^2=\{y_b^2, y_\tau^2,...\}$.
We remark that, to calculate 
these anomalous dimensions, one has to 
take into account that redundant operators removed from our operator basis are nevertheless generated through renormalization at the one-loop level. For details about how to deal with this effect, see Appendices~\ref{AppShifts} and \ref{AppRGEs}. The need to care about such effect also means that the RGEs depend on the choice of redundant operators ({\em i.e.} on the basis).


Let us  make a quantitative analysis of the size of these radiative effects. Working at one-loop leading log order,
\be
c_i(M_t)\simeq c_i(\Lambda)-\gamma_{c_i}\log\frac{\Lambda}{M_t}\ ,
\ee
which is enough if we take $\Lambda\sim  2 \text{ TeV}$ as UV scale and $M_t$ as electroweak scale, we obtain the following radiative modifications of the Wilson coefficients, $\Delta c_i \equiv c_i(M_t)-c_i(2 \text{ TeV})$:
\bea
  \Delta c_H &=& -0.17\,  c_H-0.49\, c_L^{(3)}\ ,\nonumber\\
 \Delta \lambda c_6 &=&  -0.36\, \lambda c_6-0.015\, c_H
+0.082\, c_L^{(3)}+0.244\, c_{y_t} \ , \nonumber\\
 \Delta c_{y_t} &= &-0.30\, c_{y_t}-0.035\, c_H -0.013\, c_R 
+0.043\, c_L+0.13\, c_L^{(3)}- 0.093\, c_{LR}- 0.12\, c_{LR}^{(8)}\ ,\nonumber \\
   \Delta c_{y_b}& =& -0.12\, c_{y_b}-0.068\, c_{y_t}+0.0060\, c_H
-0.012\, c_L+0.054\, c_L^{(3)}+0.027\, c^{tb}_{R}/g_*^2
\nonumber\\
& & +(0.16\, c_{y_t y_b}+ 0.027\, c^{(8)}_{y_t y_b})/g_*^2\ ,\nonumber\\
  \Delta c_{y_\tau}& =& -0.096\, c_{y_\tau} -0.081\, c_{y_t} +0.0060\, c_H + 0.16\, c_L^{(3)}\,
+(0.012 c_{y_t y_\tau} 
  +0.061 c_{y_t y_\tau}^\prime)/g_*^2 .
 \label{rgeimpact}
\eea
We see that in a few cases, the numerical impact of  operator mixing can be significant, like the mixing of $c_L^{(3)}$ into $c_H$; $\lambda c_6$ and $c_{y_t}$ into
$\lambda c_6$; and $c_{y_t}$ into itself.

\subsection{Anomalous dimensions of  constrained operators\label{RGEs2}}

Other interesting anomalous dimensions to calculate 
correspond to operators that are at present
constrained by experiments. Here we present those of $c_T$, $c_B$, $c_W$, and for the top quark, $c_R$, $c_L$, and $c_L^{(3)}$:
\bea
16\pi^2\gamma_{c_T}&=&
\frac{3}{2}{g'}^2c_H+
4N_c y_t^2(c_R-c_L)\ , 
\\[0.2cm]
16\pi^2\gamma_{c_R}&=&\left[2(4+N_c)y_t^2-9g^2-\frac{7}{3}{g'}^2\right]c_R-4(N_c+1)\left(y_t^2-\frac{2}{9}{g'}^2\right)c_{RR}
\nonumber\\
&&+2N_c\left(y_t^2+\frac{1}{9}{g'}^2\right)c_{LR}
+2y_t^2\left(\frac{1}{4}c_H-c_L\right)\ ,
\eea
\bea
16\pi^2\gamma_{c_L}&=&\left[2(2+N_c)y_t^2-9g^2-\frac{7}{3}{g'}^2\right]c_L
+2\left(y_t^2+\frac{1}{9}{g'}^2\right)\left[(2N_c+1)c_{LL}+C_Fc_{LL}^{(8)}\right]\nonumber\\
&&-2N_c\left(y_t^2-\frac{2}{9}{g'}^2\right)c_{LR}
-y_t^2\left(\frac{1}{4}c_H+c_R +9c_L^{(3)}\right)\ ,\\[0.2cm]
16\pi^2\gamma_{c_L^{(3)}}&=&
\left[2(1+N_c)y_t^2-\frac{16}{3}g^2-3{g'}^2\right]c_L^{(3)}
-2\left(y_t^2-\frac{1}{3}g^2\right)\left[c_{LL}+C_F\ c_{LL}^{(8)}\right]\nonumber\\
&&+y_t^2\left(\frac{1}{4}c_H-3c_L\right),
\\[0.2cm]
16\pi^2\gamma_{c_W }&=&  \frac{1}{3}g_*^2\left[16 N_c c_L^{(3)}-c_H\right] , 
\\[0.2cm]
16\pi^2\gamma_{c_B }&=& \frac{1}{3}g_*^2 \left[\frac{8}{3} N_c \left(2 c_R + c_L  \right)-c_H \right] \ .
\label{rge2}
\eea
From them 
we can calculate 
the  leading-log corrections to   
$c_B+c_W$, $c_T$ and $c_L+c_L^{(3)}$
that  are highly constrained  by  $\widehat S$, $\widehat T$ and the $Zbb$-coupling, as has been discussed in Section~\ref{sec:expconst}. In this way,
 coefficients that are more loosely constrained by direct processes, such as $c_H$, $c_L$ or $c_R$,
can get indirect bounds from LEP1 and Tevatron measurements.

Integrating the RGEs of \eq{rge2},  
at the one-loop leading-log order,  between the cutoff scale $\Lambda=2$ TeV and the electroweak scale,
that we take here $M_t$, one gets~\footnote{
The effects of $c_H$  and 
those of $c_{L,R}$ on $\widehat T$ were  already calculated in \cite{Barbieri:2007bh} and  \cite{Pomarol:2008bh} respectively.}
\bea
\label{ewpc1}
\Delta \widehat T&=& \Delta c_T\xi = \left[-0.003\, c_H+0.16\, ( c_L- c_R )\right]\xi\ ,\\
\Delta \widehat S& =&\Delta (c_B + c_W)\frac{M^2_W}{\Lambda^2}
=\left[  0.001\, c_H -0.01\, c_R -0.004\, c_L -0.03\, c_L^{(3)}\right]\xi\ ,\\
\Delta\frac{\delta g_{Z}^{b_L}}{g_{Z}^{b_L}}&=& \frac{\Delta [c_L+ c_L^{(3)}]}{1-(2/3)\sin^2\theta_W}\xi\simeq  \Delta [c_L+ c_L^{(3)}]\xi \nonumber\\
&= &
\left[ 0.01\, c_R-0.03\, c_L + 0.06\, c_L^{(3)}-0.17\, c_{LL} -0.0064\, c_{LL}^{(8)}+0.08\, c_{LR}\right]\xi\, ,
\label{ewpc3}
\eea
where $\Delta c_i \equiv c_i(M_t) - c_i(2 \text{ TeV})$.  Notice that even if  a $P_{LR}$ symmetry  of the BSM sector enforces $c_L+ c_L^{(3)}=0$, we can  have a nonzero $c_L+ c_L^{(3)}$ from 
the  RG running,  since the SM does not respect this parity.
The fact that the three quantities above are constrained at the per-mille level implies that the top coefficients, $\{c_L, c_R, \dots\} \times\, \xi$ cannot be of order one.
Obviously, we are barring the possibility of cancellations between the initial value of the Wilson coefficients at the scale $\Lambda$ and the radiative effects $\sim\gamma_{c_i}\log(\Lambda/M_t)$, that could only be possible by  accident.

\section{RGE impact  on   the predictions of Wilson coefficients}
\label{one-loop-examples}

Here we want to  study the  impact of the evolution of the Wilson coefficients 
from the UV scale $\Lambda$ down to the electroweak scale at which they affect Higgs physics.
This running can modify the predictions arising from  BSM models.
We  present three examples:
two-Higgs doublet models (2HDM), 
universal theories, and  scenarios with sizeable $c_{L,R}$,  such as  composite-top models.
\newline

\noindent{\bf 2HDM and Supersymmetric theories:}
At tree-level, assuming ordinary $R$-parity, the only $d=6$ operators that can be induced
in supersymmetric models arise  from the exchange of the extra Higgses since these are  the only $R$-even heavy fields.
In particular,  the MSSM
contains  an extra heavy Higgs doublet.
It is therefore well motivated to look for the impact  of an extra heavy Higgs doublet
in SM Higgs physics.

Denoting the heavy Higgs by $H'$, defined to have $Y_{H'}=1/2$, its relevant couplings to the SM fermions and Higgs are given by
\be
{\cal L}' =  -\alpha_u  y_u  \bar Q_L \widetilde H' u_R
-\alpha_d     y_b \bar Q_L H' d_R -\alpha_e   y_e \bar l_L H'	e_R -  \lambda'   H^{\prime\dagger}  H |H|^2+h.c. + \cdots\  ,
\ee
where $\alpha_{u,d,e}$ are constants and we assume that $\lambda'$ is a real number. 
In particular 2HDMs, these constants are 
\bea
&&\alpha_u=\alpha_d=\alpha_e=\tan\beta\ ,\qquad   \quad \qquad \ \ \ \  \ \text{for type-I 2HDM}\\
&&\alpha_u=-\cot\beta\ ,\ \ \alpha_d=\alpha_e=\tan\beta\ ,\qquad   \text{for type-II 2HDM (MSSM)}
\eea
where $\tan\beta$ defines the rotation from the original basis, in which
only one Higgs couples to a given type of fermion, to the mass-eigenstate basis
before EWSB. 
At the order we work ($\sim v^2/\Lambda^2$), $\tan\beta$ coincides with that defined in the MSSM.
Integrating out this doublet at tree-level, we obtain 
the following nonzero coefficients for the third-family $d=6$ operators:
\be
\begin{split}
& g^2_*c_{y_t}=  \alpha_t \lambda^{\prime} \ ,\ \ \
g^2_*c_{y_b}= \alpha_b \lambda^{\prime} \ ,\ \ \
g^2_*c_{y_\tau}= \alpha_\tau \lambda^{\prime} \ , \ \ \ 
g_*^2  \lambda c_6 = \lambda^{\prime 2 } \ ,  \\
& g^2_*c_{LR}^{(8)}=2 N_c g^2_* c_{LR}=- \alpha_t^2 y^2_t  \ ,\ \ \
c_{y_ty_b}= \alpha_t\alpha_b\  ,\ \ \
c_{y_t y_\tau} =\alpha_t \alpha_\tau \  .
\end{split}
\ee
We have used $
(\bar Q_L t_R)(\bar t_R Q_L)= - (\bar Q_L T^A \gamma^\mu Q_L) (\bar t_R T^A \gamma_\mu t_R) -  (\bar Q_L  \gamma^\mu Q_L) (\bar t_R  \gamma_\mu t_R)/(2N_c)$ and now  $\Lambda=M_{H'}$.
Under the RGE flow of Eqs.~(\ref{ct})-(\ref{ctau})
 the operators ${\cal O}_{y_f}$ mix with ${\cal O}_{LR}, {\cal O}_{LR}^{(8)}$, ${\cal O}_{y_uy_d}$ and ${\cal O}_{y_uy_\tau}$.
In the  type-II 2HDM, we obtain in the one-loop leading-log approximation and neglecting $O(\lambda,g^2,{g'}^2)$ corrections:
\bea
g^2_*c_{y_t}(m_h)&=&  -\frac{\lambda^{\prime}}{t_\beta}\left[1-\frac{21y^2_t }{16\pi^2}\log\frac{M_{H'}}{m_h}\right]
+\frac{3y^4_t}{4\pi^2 t^2_\beta}\log\frac{M_{H'}}{m_h}\nonumber \ ,\\
g^2_*c_{y_b}(m_h)&=&  \lambda^{\prime }t_\beta\left[1- \frac{y^2_t}{2\pi^2} \log\frac{M_{H'}}{m_h}\right]
+\frac{y^2_t}{16\pi^2}\left[5\frac{\lambda^{\prime }}{t_\beta}
- 14 y_t^2\right]\log\frac{M_{H'}}{m_h} \ ,\nonumber\\
g^2_*c_{y_\tau}(m_h)&=&  \lambda^{\prime }t_\beta\left[1 - \frac{3y_t^2}{8 \pi^2}   \log\frac{M_{H'}}{m_h} \right] + \frac{ 3y_t^2}{8 \pi^2} \left[\frac{\lambda^{\prime}}{t_\beta}- 2 y_t^2  \right] \log\frac{M_{H'}}{m_h}
\ ,\label{ghff}
\eea
with $t_\beta\equiv \tan\beta$. 

\begin{figure}[t]
$$\includegraphics[width=0.47\textwidth]{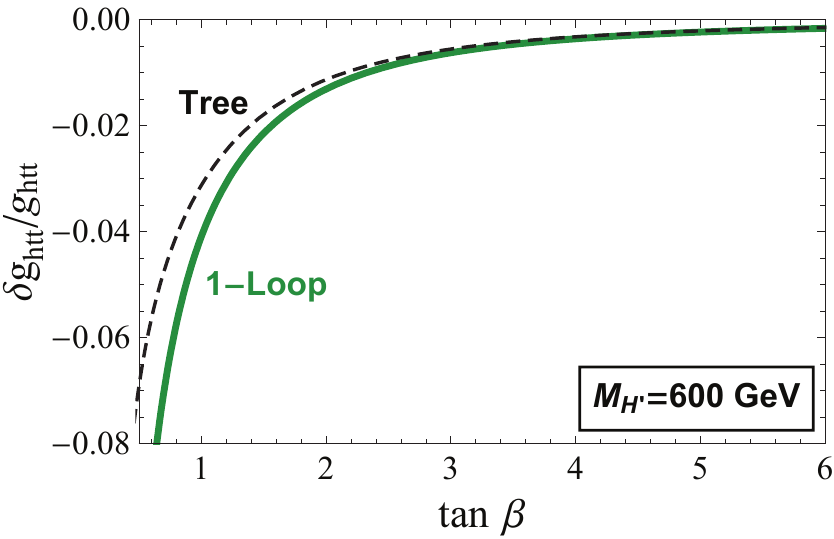}    \qquad
\includegraphics[width=0.46\textwidth]{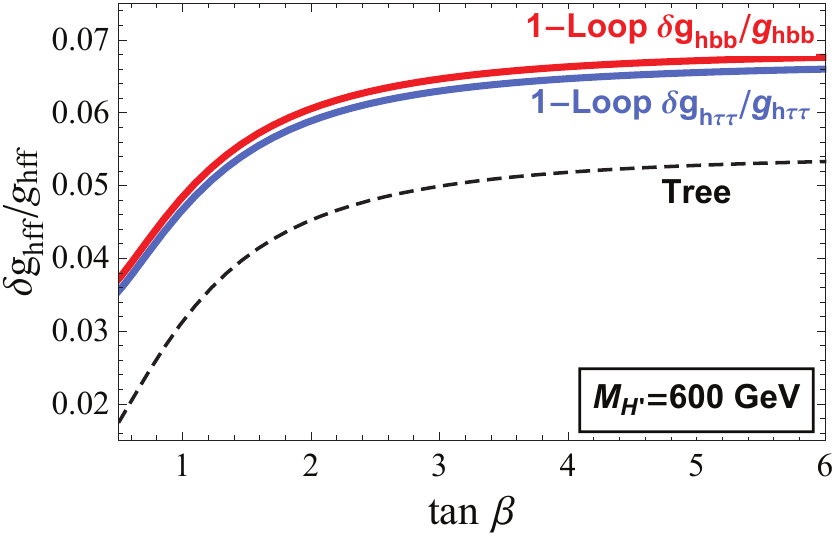}$$
\begin{center}
\caption{\emph{Relative modification of the Higgs coupling to fermions, $\delta g_{hff}/g_{hff}=-c_{y_f}\xi$, 
Eq.~(\ref{hff}), at tree-level (dashed line) and after including RGE effects from $\Lambda$ to the electroweak scale (solid lines) as a function of $\tan\beta$ in an MSSM scenario with $\Lambda=M_{H'}=600$ GeV and unmixed stops
heavy enough to reproduce $m_h=125$ GeV.  Left plot: top coupling.  Right plot:  bottom (lower solid line) and tau (upper solid line) couplings.}}
 \end{center}
\end{figure}

\begin{figure}[t]
$$\includegraphics[width=0.6\textwidth]{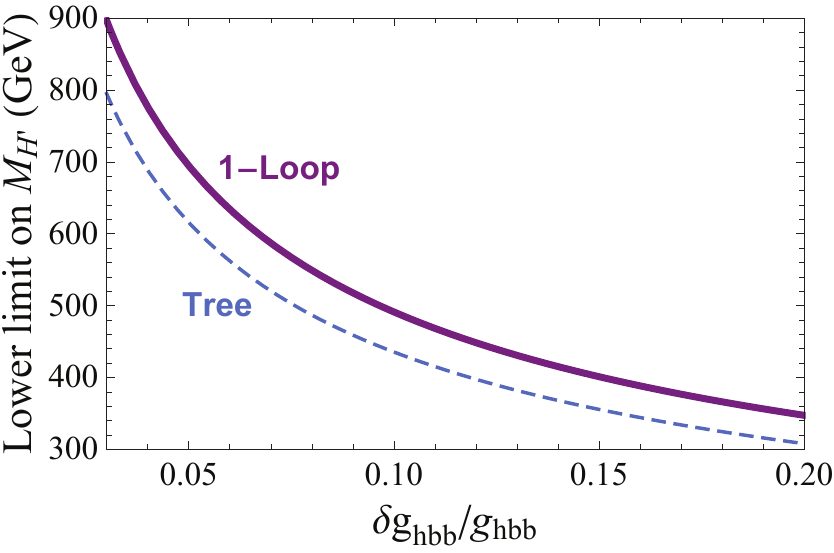}$$
\begin{center}
\caption{\label{RGimpact}
\emph{ Lower bound on $M_{H'}$ as a function of the 
upper bound on the relative deviation $\delta g_{hbb}/g_{hbb}$,
in an MSSM scenario with $\tan\beta=5$ and unmixed stops
heavy enough to reproduce $m_h=125$ GeV. The dashed line corresponds to a tree-level analysis (parameters calculated at the scale $M_{H'}$), while the solid line includes the RG running from $M_{H'}$ down to $m_h$.
}}
 \end{center}
\end{figure}

To illustrate the impact of these radiative effects, let us  consider the MSSM, a model 
 which predicts  $\lambda^{\prime}=(1/8)(g^2+{g'}^2)\sin 4 \beta$ at tree-level \cite{gmr}. 
We take  the stop mass scale  $M_{\tilde t}$ large enough to get $m_h\simeq 125$ GeV through the well-known loop
corrections to the Higgs quartic coupling, which at one-loop and zero stop mixing read:
\be
\lambda(m_h)=\frac{1}{8} (g^2+{g'}^2)\cos^22\beta+
\frac{3y_t^4}{16\pi^2}\log\frac{M_{\tilde t}^2}{M_t^2}\ ,
\ee
which is precise enough for our illustrative purposes. For consistency we must also include similar radiative corrections to $\lambda'$, which read at one-loop:
\be
\lambda'(M_{H'})=\frac{1}{8} (g^2+{g'}^2)\sin 4\beta-
\frac{3y_t^4}{8\pi^2t_\beta}\log\frac{M_{\tilde t}^2}{M_{H'}^2}\ .
\ee
This gives the value of $\lambda'$ that we can then plug in Eq.~(\ref{ghff})
to obtain the  RG-improved corrections for  $g_{hff}$  induced by integrating out the heavy Higgses. The result is shown 
as a function of $t_\beta$ in Fig.~1, which compares the tree-level result (dashed lines) and the one-loop result (solid lines) which takes into account the running from $\Lambda=M_{H'}$ down to the electroweak scale $m_h$. One sees that the effect of the running can be quite significant, easily $\sim 50\%$ or more.
The importance of this effect can be further appreciated in Fig.~2,
which shows the lower bound one could set on $M_{H'}$ from an upper
bound on $\delta g_{hbb}/g_{hbb}$, the deviation of  $g_{hbb}$  from its SM value.
By comparing the tree-level bound (dashed line) and the one-loop bound (solid line) one sees that the bound is shifted significantly by the inclusion of the RG corrections from $M_{H'}$ to $m_h$.

Finally, notice that $c_H$, which is not generated in the MSSM at tree-level since there are no heavy $R$-even singlet states,
 is not generated by  the RGE evolution and therefore is also zero in the leading-log approximation.
\newline

\noindent{\bf Universal theories and composite Higgs models:}  Universal  theories predict
 $c_{y_u}=c_{y_d}=c_{y_e}$.
This prediction
is modified by the evolution of these coefficients from the scale $\Lambda$, where they are generated, down to the electroweak scale.
In particular,  for $\Lambda=2$ TeV, we find that the 
breaking of universality due to the top Yukawa coupling gives 
\be
\begin{split}
& c_{y_t}(m_h)=c_{y_b}(m_h)\left(1-\frac{8y^2_t}{16\pi^2}\log\frac{\Lambda}{m_h}\right)-
\frac{3y^2_tc_H}{16\pi^2}\log\frac{\Lambda}{m_h}
\simeq 0.88 c_{y_b}(m_h)- 0.05c_H\ , \\
&  c_{y_b}(m_h)=c_{y_\tau}(m_h)\left(1-\frac{y^2_t}{16\pi^2}\log\frac{\Lambda}{m_h}\right)
\simeq 0.98 c_{y_\tau}(m_h) \ .
\end{split}
\ee
This is a sizeable departure from universality for $c_{y_t}$ that will have  to be taken into account when fitting these models to data.
Also it is worth noticing  that in  models in which    only $c_H$ is generated (models with only heavy singlets) and $c_{y_f}(\Lambda)=0$,
the value  of $c_{y_f}$ is also very small at low-energies,  $c_{y_f}(m_h)\simeq 0$.
In the minimal composite Higgs model, we also have the prediction $c_H=1$ at $\Lambda\sim$ 2 TeV \cite{Giudice:2007fh}.
We find that the RG effects  give a $\sim 20\%$ reduction of this prediction.
\newline

\noindent{\bf  Models with a non-SM top:}
The top is the only quark whose properties are not yet measured at high precision, allowing then sizeable deviations
from their SM predictions. 
There are also theoretical motivations
to expect the top to be the quark
with the largest deviations from the SM predictions, as it is the quark with the largest 
coupling to the Higgs.
This is specially true in composite Higgs models
where one expects the top to show also certain  degree of compositeness. 
In these examples we can expect sizable values  for $c_R$, 
$c_L^{(3)}$, $c_L$ and $c_{LR}$ that can affect, at the one loop-level, the Higgs coefficients $c_H$ and $c_{y_f}$.
As it is clear from \eq{rgeimpact}
the effects of $c_R$ on the RGE evolution of $c_H$ and $c_{y_f}$ are very small. 
Nevertheless,  those from $c_L^{(3)}$ and $c_L$
are quite sizeable, even in the limit   $c_L\simeq -c_L^{(3)}$ as  required in order  to avoid large  tree-level contributions to $Zb\bar b$.
Unfortunately these coefficients also give large one-loop effects to the $\widehat T$ and $\widehat S$ parameters and $Zbb$, as     Eqs.~(\ref{ewpc1})-(\ref{ewpc3}) show,
    that bounds them to be small (unless $\xi$ is small). 
   Interestingly, the coefficient $c_{LR}^{(8)}$ is not constrained by   Eqs.~(\ref{ewpc1})-(\ref{ewpc3}).
Therefore it can give sizeable contributions to the RGE evolution of $c_{y_t}$:
\be
c_{y_t}(m_h)=c_{y_t}(\Lambda)-\frac{2 y^2_t}{3\pi^2}c_{LR}^{(8)}\log\frac{\Lambda}{m_h}\, ,
\ee
that  is of order    $\sim 15\%$.
A nonzero $c_{LR}^{(8)}$ could arise from integrating out a massive gluon coupled to the  top.

\section{Conclusions}

As the measurements of the  Higgs properties improve,
it will be important to  understand their implications for BSM models.
In this article we have adopted the framework of effective Lagrangians as a tool to study
the effects of $d=6$ operators in Higgs physics.
  As a first step, we have discussed the choice of basis of  operators. 
Our basis has been defined    following \cite{Giudice:2007fh,us} that  
    distinguished two classes of operators: tree-level (or current-current) operators, and one-loop operators.
This choice can be important 
when calculating one-loop operator mixing,
since most of the  tree-level operators do not mix with  one-loop operators  under RG evolution \cite{us}.
 Another important  property of our basis is that it contains a subset of 5 CP-even operators made of Higgs and gauge field-strengths,  that in our case are  ${\cal O}_{W,B}$  ${\cal O}_{HW,HB}$ and ${\cal O}_{BB}$, (leaving aside  ${\cal O}_{GG}$).  
We have found that it  is important to keep these 5 operators to make the connection with experiments more 
transparent  [these subset could  also be written with   ${\cal O}_{WB,WW}$ by using the identities (\ref{OpId1}) and (\ref{OpId2})].
Bases, such as \cite{Grzadkowski:2010es} and \cite{concha}, that eliminate    two of these operators
in favor of   operators made of SM fermions, as it  can be done by using the EoM, have   dangerous blind directions for LEP1 experiments, 
which
make the   contact with experiments more difficult.

We have calculated  the modifications that the operators of the effective Lagrangian induce in the Higgs couplings relevant for the main decays and production mechanisms.
It has been shown that these operators can be divided in two subsets.
There are 11 operators (for one family) with coefficients given in \eq{relevantcis}, 
 that can only affect Higgs physics  and no other SM processes at tree-level.
The  number 11 can be deduced from  counting the number of independent operators one can write as
$|H|^2 {\cal O}_4$ with  ${\cal O}_4$ a $d=4$ operator formed with SM fields.
The second subset, formed by the rest of  operators,  
enter in other SM processes and therefore can be constrained
by non-Higgs experiments.
Among the latter, considering only the CP-even ones, 
we have found that the least constrained   correspond to
 the two combinations of Wilson coefficients 
appearing in  the measurements of the $ZWW/\gamma WW$ coupling, \eq{3vert},
 that LEP2 has only constrained at the few per-cent level.
  LHC will probe these vertices with better accuracy,  so that it will be able to improve these constraints.

We have calculated the anomalous dimensions of the 5 tree-level operators of the list \eq{relevantcis},
which allows us to calculate the running of the coefficients from the high-energy scale $\Lambda$
where they are generated down to the electroweak scale.
All technical details of these calculations have been discussed
 in Appendix~\ref{AppRGEs}.
Since  $\widehat S$ and $\widehat T$ parameters, and the $Zb\bar b$ coupling are very well constrained, 
we have also calculated the anomalous dimension of the operators contributing to these quantities. 
In this way, we can put indirect bounds on Higgs operators.

We have applied our results to BSM models as MSSM, universal theories (as composite Higgs models)
and models with non-standard top couplings.
In such models we have evaluated
 the leading-log corrections to the predictions for the Higgs couplings.
The corrections from the running can be quite large for $\Lambda\sim$ few TeV, 
 as Fig.~\ref{RGimpact} shows.
Our calculation of the anomalous dimensions is an aspect of the physics of the $d=6$ operators which will become more relevant as soon as we have better measurements of the Higgs couplings.

\section*{Acknowledgements}
We thank M. Carrel, C. Grojean, M. Montull, F. Riva and A. Wulzer
for useful discussions.
J.R.E. thanks CERN for hospitality and partial financial support during some stages of this work.
This work has been partly supported
by Spanish Consolider Ingenio 2010 Programme CPAN (CSD2007-00042) and
the Spanish Ministry MICNN under grants FPA2010-17747 and
FPA2011-25948; and the Generalitat de Catalunya grant 2009SGR894.
The work of A.P. has also been  supported by  the ICREA Academia Program. The work of J.E.M. has been supported by the Spanish Ministry MECD through the FPU grant AP2010-3193.

\appendix
   \numberwithin{equation}{section}

\newpage
\section{Currents, redundant operators and field shifts} 
\label{AppShifts}

In this Appendix we first list, in Subsection~\ref{A.currents}, the different currents (of dimension $\leq 3$) built from SM fields
that enter into the $d=6$ current-current operators.  We 
examine in Subsection~\ref{A.currentcurrent} how these operators can be generated from integrating out heavy particles discussing what type of operators appear depending on the quantum numbers of the heavy fields. Some of these operators are redundant and can be eliminated from the Lagrangian by using the field equations of motion or, equivalently, by field redefinitions. We discuss this point in Subsection~\ref{A.shifts}, where we  give a possible set of  field redefinitions that can be used to get rid of the redundant operators. 

\subsection{Currents of SM fields\label{A.currents}}
For simplicity we limit our examples of currents to the SM with a single family of fermions, the generalization to 3 families being straightforward. The scalar currents are:
\be
\begin{array}{lll}
 J_H= |H|^2\ , &
J_H^{(2)}=H|H|^2\  , & 
J_{\Box H} = D_\mu^2 H\ , \\
J_H^{a}=H^\dagger\sigma^a H\ , &
J_{y_f H}=y_f \bar F_L f_R\ , &
 J_{y_f H}^{A}=y_f \bar Q_L T^A f_R\ , 
\end{array}
\ee
where $T^A$ are the $SU(3)_c$ generators and from now on we use the notation $F_L=\{Q_L,L_L\}$ and $f_R=\{u_R,d_R,e_R\}$ for fields,  while $F=\{q,l\}$ and $f=\{u,d,e\}$ are used for the corresponding operator indices.
Obviously, one can also have the conjugate currents: $ \tilde{J}_{H}^{(2)}=\tilde H |H|^2$, $\tilde{J}_{\Box H}=D_\mu^2 \tilde H $, etc. 

There are also
vector currents made of SM bosons, like:
\be
\begin{array}{lll}
 J_H^\mu=i H^\dagger\lra{D^\mu} H\ , &
J_{W_R}^{\mu}= i \widetilde{H}^\dagger\lra{D^\mu} H\ , & 
J_H^{a\, \mu}=i H^\dagger\sigma^a  \lra{D^\mu} H\ , \\
J_{ B}^{\mu}= \partial_\nu B^{\mu\nu}\ , & 
 J_W^{a\, \mu}= D_\nu W^{a\, \mu\nu}\, ,&
J_G^{A\, \mu}= D_\nu G^{A\, \mu\nu}\ ,
\end{array}
\ee
and  made of SM fermions, like:
\be
\begin{array}{lll}
J_{f f^\prime}^{\mu}=\bar f_R\gamma^\mu f^\prime_R\ , &
J_F^{\mu}=\bar F_L\gamma^\mu F_L\ ,  &  
J_F^{a\, \mu}=\bar F_L\sigma^a\gamma^\mu F_L\ ,\\
J_{f }^{A\, \mu }=\bar f_R T^A \gamma^\mu f_R\ ,&
J_Q^{A\, \mu}=\bar Q_L T^A \gamma^\mu Q_L\, ,&  
\end{array}
\ee
as well as the lepto-quark currents:
\be
J^\alpha_{Q e} = y_e \bar Q_L^\alpha e_R\ ,\quad
J^\alpha_{L u} = y_u \bar L_L u^\alpha_R\ ,\quad
J^\alpha_{L d} = y_d \bar L_L d^\alpha_R\ ,
\ee
where we write explicitly the color index $\alpha$. Finally, we list fermionic currents made of SM fields. They can be $SU(2)_L$ singlets:
\be
J_{Df}=i \Dslash f_R\ , \ \ \ 
J_{y_f\bar f_R}= \{ y_u^\dagger  \widetilde{H}^{\dagger}Q_L\ ,  
y_d^\dagger  H^{\dagger}Q_L\ , y_e^\dagger H^\dagger L_L\} \ ,
\label{ferC1}
\ee
doublets:
\be
 J_{DF}=i \Dslash F_L \ , \ \ \ 
J_{y_f\bar F_L}= \{ y_u  \widetilde{H} u_R\ ,   
y_d H d_R\ , y_e H e_R\}\ , 
\label{ferC2}
\ee
or triplets:
\be 
 J^{a}_{\widetilde{H}F}= \{\widetilde{H}^\dagger \sigma^a Q_L ,
\widetilde{H}^\dagger \sigma^a L_L\}\ , \ \ \ 
J^a_{HF}=\{H^\dagger  \sigma^a Q_L , H^\dagger  \sigma^a L_L \}\ . \ \ \  
   \label{ferC3}
\ee
The previous list of SM currents is not complete but contains all the currents necessary to build the current-current operators of our basis (defined in the main body of the paper), as well as many of the redundant operators.

\subsection{Current-current operators\label{A.currentcurrent}}

The $d=6$ current-current operators can in principle be generated from the tree-level exchange of heavy fields. We can then classify such operators by the quantum numbers of the exchanged heavy fields.  We present such classification below (giving explicit expressions for those redundant operators that appear here for the first time.). Finding possible deformations in SM couplings that can be assigned to particular current-current operators can offer crucial information in identifying the 
heavy physics responsible for such effects.

\begin{enumerate}
\item[$\circ$]  {\bf Scalar $\times$ scalar}

The exchange of a heavy scalar singlet can lead (after integration by parts) to:
\be
-J_H \Box J_H=2{\cal O}_H \ .
\ee
From  a heavy  scalar $SU(2)_L$-doublet   we get:
\be 
\begin{split}
& \lambda J_H^{(2)\, \dagger} J_H^{(2)}= {\cal O}_6\ , \quad\quad\quad\quad\quad
J_H^{(2)\, \dagger}  J_{\Box H} + \text{h.c.}= - 2 (\mathcal{O}_H+\mathcal{O}_r)\ , \ \\ 
&
 J_{y_uH} \tilde{J}_{H}^{(2)}  = {\cal O}_{y_u} \ , \quad\quad\quad\quad\quad
 J^\dagger_{y_d H}J_{y_e H} = {\cal O}_{y_dy_e}
\ , \\ 
& (J_{y_u H})^r\epsilon_{rs}(J_{y_e H})^s= {\cal O}_{y_uy_e}\ ,\quad  (  J_{y_uH} )^r \epsilon_{rs} (   J_{y_d H})^s = 
 \mathcal{O}_{y_u y_d} \ , \ \
\end{split}
\ee
and also:
\bea
&& J^\dagger_{\Box H}  J_{\Box H}  = \left| D_\mu^2 H \right|^2 \equiv {\cal O}_{K4}\ , \nonumber\\
&& J_{y_u H} \tilde{J}_{\Box H} = - y_u D_\mu \left( \bar{Q}_L u_R \right) D^\mu \widetilde{H} \equiv -{\cal O}^u_{yH}\ .
\eea

If the heavy doublet is also charged under $SU(3)_c$ we can get:
\be
(  J_{Qe}^\alpha )^r \epsilon_{rs} (   J^\alpha_{Lu})^s =  \mathcal{O}_{y_u y_e}^{\prime} \ ,\quad
(   J^A_{y_u H} )^r  \epsilon_{rs} (  J^A_{y_d H})^s =  \mathcal{O}^{(8)}_{y_u y_d} \ ,
\ee
while, from a heavy  scalar $SU(2)_L$-triplet   we would obtain:
\be
J_H^{a}  D^2 J_H^{a}=- 2\mathcal{O}_T - 4 \mathcal{O}_r\ .
\ee

\item[$\circ$] {\bf Vector $\times$ vector}

From the exchange of a heavy  singlet vector one can get:
\be
\begin{split}
& J^\mu_H J_{H\, \mu}=-2{\cal O}_T\ , \ \ \  
 g^{\prime} J_{B}^\mu J_{H \mu}=  2{\cal O}_B\ , \ \ \ 
 J_{uu}^{\mu} J_{H \, \mu} = {\cal O}_{R}^u \ ,    \\
&  J^\mu_H J_{F\, \mu}={\cal O}_{L}^F\ , \ \ \
J^{\mu}_B J_{B\, \mu}=-2{\cal O}_{2B}\ , \ \ \ 
J_{uu}^{\mu}  J_{uu \, \mu}  = {\cal O}_{RR}^u \ , \ \\  
& J_{uu}^{\mu} J_{F \, \mu} = {\cal O}_{LR}^u \ , \ \ \  
 J^\mu_F J_{F\, \mu}={\cal O}_{LL}^F\ , \ \ \  
y_u^\dagger y_d J_{W_R\,\mu} J^{\mu}_{ud} = {\cal O}^{ud}_R
\ ,
\end{split}
\ee
as well as
\bea
& g^{\prime} J^{\mu}_B J_{uu\, \mu}=g'(\bar u_R \gamma^\mu u_R)(\partial^\nu B_{\mu\nu})\equiv {\cal O}^u_{BR}\ , \nonumber\\
& g^{\prime} J^{\mu}_B J_{F\, \mu}=g'(\bar F_L \gamma^\mu F_L)(\partial^\nu B_{\mu\nu})\equiv {\cal O}^F_{BL}\ .
\eea

The exchange of a heavy  $SU(2)_L$-triplet vector can produce:
\be
\begin{split}
& J^{a\, \mu}_H J_{H\, \mu}^a=-2{\cal O}_H+4{\cal O}_r\ , \ \ \ 
g J_{W\, \mu}^a J^{a\, \mu}_H =2 {\cal O}_W\ , \ \ \ 
  J_F^{a\, \mu} J_{H \, \mu}^a = {\cal O}_{L}^{(3)F} \ , \ \ \ \\
& J^{a\, \mu}_W J_{W\, \mu}^a=-2 {\cal O}_{2W}\ , \ \ \ 
J_{L\, \mu}^a  J_L^{a\, \mu}= {\cal O}_{LL}^l\ , \ \ \
 J_{Q\, \mu}^a  J_Q^{a\, \mu}= 4 {\cal O}_{LL}^{(8)q}+\frac{2-N_c}{N_c} {\cal O}_{LL}^q\ , 
\end{split}
\ee
and
\be
 g J^{a\mu}_W J_{F\, \mu}^a=g(\bar F_L \gamma^\mu \sigma^a F_L)(D^\nu W^a_{\mu\nu})\equiv {\cal O}^F_{WL}\ ,
\ee
while a heavy  $SU(3)_c$-octet vector could give:
\be
 J_{u \, \mu}^A  J_u^{A\, \mu}= (1/3)\ {\cal O }_{RR}^u\ , \ \ \
  J_{Q \,  \mu}^A  J_Q^{A\, \mu}={\cal O}_{LL}^{(8)\, q}\ , \ \ \
   J_{Q \,  \mu}^A  J_u^{A\, \mu}={\cal O}_{LR}^{(8)\, u}\ . \ \ \
   \ee

\item[$\circ$] {\bf Fermion $\times$ fermion}

Finally, we list operators that can arise from integrating a heavy fermion. If the fermion is a singlet:
\be
\begin{split}
\bar J_{y_u \bar u_R} i \slashed{D} J_{Du} & = y_u D_\mu(\bar Q_L \widetilde H)\gamma^\mu\gamma^\nu D_\nu u_R \equiv  \mathcal{O}^u_{yR} \ , \ \ \  \\
\bar J_{y_u \bar u_R} i \slashed{D} J_{y_u \bar u_R} + \text{h.c. }  & = \frac{1}{2} |y_u|^2 \left[ -\widetilde{\cal O}_L^{(3)q}+\widetilde{\cal O}^q_L-{\cal O}_L^{(3)q}+{\cal O}^q_L \right] \ , \ \ \  \\
\end{split}
   \label{ferO1}
\ee
where
\bea
 \widetilde{\cal O}_L^{(3)q}&=& i (\bar Q_L \sigma^a\lra\Dslash Q_L)(H^\dagger \sigma^a H)\ ,\nonumber\\
\widetilde{\cal O}^q_L&=& i ( \bar Q_L \lra\Dslash Q_L)|H|^2\ ,
\eea
are redundant operators.

If the fermion integrated-out is a doublet, one can get:
\be 
\begin{split}
\bar J_{DQ} i \slashed{D} J_{y_u\bar Q_L}  &= y_u D_\mu\bar Q_L\gamma^\mu\gamma^\nu D_\nu( \widetilde H u_R) \equiv \mathcal{O}^u_{yL} \ , \ \ \  \\
\bar J_{y_u\bar Q_L} i \slashed{D} J_{y_u\bar Q_L} + \text{h.c. } & =   |y_u|^2 \left[- {\cal O}^u_R+\widetilde{\cal O}^u_R \right] \  , \ \ \  \\
\end{split}
   \label{ferO2}
\ee
with the redundant operator:
\be
\widetilde{\cal O}^u_R=  i ( \bar u_R \lra\Dslash u_R)|H|^2\ .
\ee 
Finally, from integrating out a heavy  fermion triplet, we can get: 
\be
\begin{split} 
\bar  J^a_{\widetilde{H}F} i \slashed{D} J^{a}_{\widetilde{H}F} +\mathrm{h.c.} &=\frac{1}{2}  \left[ \widetilde{\cal O}_L^{(3)F}+3\widetilde{\cal O}^F_L+{\cal O}_L^{(3)F}+3{\cal O}^F_L \right] \ , \ \ \ \\
\bar J^a_{HF} i \slashed{D} J^a_{HF} +\mathrm{h.c.} & =\frac{1}{2}  \left[- \widetilde{\cal O}_L^{(3)F}+3\widetilde{\cal O}^F_L+{\cal O}_L^{(3)F}-3{\cal O}^F_L \right] \ . \ \ \  \\
\end{split}
     \label{ferO3}
\ee

To describe the effect of a heavy fermion that is a color octet, one would need to generalize the quark currents of Subsection~\ref{A.currents} by inserting $SU(3)_c$ generators.  However, the dimension-6 operators that result have been already found in Eqs.~(\ref{ferO1}) and (\ref{ferO2}).

\end{enumerate}

\subsection{Field redefinitions and redundant operators\label{A.shifts}}
Many $d=6$ current-current operators are redundant: they can be removed from the Lagrangian by field redefinitions. We will show how field redefinitions can be used for that purpose, focusing here on current-current operators not of the 4-fermion type.

Let us start first with bosonic operators. Consider the
following  transformations that shift fields by some of the bosonic currents listed in Subsection~\ref{A.currents} (with the same quantum numbers of the shifted fields):
\bea
H\rightarrow H +\alpha_1  J^{(2)}_H/\Lambda^2\ ,&&
H\rightarrow H\left(1-\alpha_2  m^2/
\Lambda^2\right)+\alpha_2 J_{\Box H}/\Lambda^2\ ,\nonumber\\
B_\mu \rightarrow B_\mu + [ g'\alpha_B  J_{H\, \mu}+ \alpha_{2B}  J_{B\, \mu}
]/\Lambda^2\ ,&&
W^a_\mu \rightarrow W^a_\mu +[ g \alpha_W 
J_{H\, \mu}^a + \alpha_{2W}  J_{W\, \mu}^a] /\Lambda^2\ ,\nonumber\\
G^A_\mu \rightarrow G^A_\mu + \alpha_{2G} J^A_{G\, \mu}/\Lambda^2\ ,&&
\label{fieldredef}
\eea
with $\alpha_i$ arbitrary parameters (taken real). These transformations  induce shifts in the $d=6$  Wilson coefficients \footnote{Shifts of 
order $m^2/\Lambda^2$ induced on the renormalizable dimension-4 SM operators play no role. There are also shifts in the coefficients of the operators made of fermions that we show below.} of  Eqs.~(\ref{first6dim}) and (\ref{second6dim}) plus the redundant operator ${\cal O}_{K4}=|D_\mu^2 H|^2$:
\bea
&&c_H\rightarrow c_H +2\alpha_1+(4\lambda\alpha_2-\alpha_W g^2)/g_*^2\ ,\nonumber\\
&&c_r\rightarrow c_r +2\alpha_1+(4\lambda\alpha_2+2\alpha_W g^2)/g_*^2\ ,\nonumber\\
&&c_6\rightarrow c_6-4\alpha_1\ ,\nonumber\\
&&c_T\rightarrow c_T- \alpha_B {g'}^2/g_*^2\ ,\nonumber\\
&&c_B\rightarrow c_B-2\alpha_B+\alpha_{2B}\ ,\nonumber\\
&&c_W\rightarrow c_W-2\alpha_W+\alpha_{2W}\ ,\nonumber\\
&&c_{2W}\rightarrow c_{2W}+2\alpha_{2W}\ ,\nonumber\\
&&c_{2B}\rightarrow c_{2B}+2\alpha_{2B}\ ,\nonumber\\
&&c_{2G}\rightarrow c_{2G}+2\alpha_{2G}\ ,\nonumber\\
&&c_{K4}\rightarrow c_{K4}-2\alpha_2\, .
\eea
Notice that only operators of tree-level type are shifted. 
Using this shift freedom, we could eliminate 7 out of the 10 operators $\{
{\cal O}_H,{\cal O}_r,{\cal O}_6,{\cal O}_T,{\cal O}_B,{\cal O}_W,{\cal O}_{2W},{\cal O}_{2B},{\cal O}_{K4}, {\cal O}_{2G}
\}$ by choosing appropriately the $\alpha_i$'s and leave only ${\cal O}_H$, ${\cal O}_T$ and ${\cal O}_6$. As we discussed in Section~\ref{basis}, however, it is convenient to keep the operators ${\cal O}_W$ and ${\cal O}_B$ in the basis, in which we could also keep ${\cal O}_{2W}$, ${\cal O}_{2B}$ and ${\cal O}_{2G}$. If we do not use  5 of these shifts to remove ${\cal O}_W$, ${\cal O}_B$, ${\cal O}_{2W}$, ${\cal O}_{2B}$ and ${\cal O}_{2G}$,  they can be used later on   to remove 5 other operators involving fermions. We will discuss such operators next.

Besides the bosonic redundant operators discussed above, there are redundant operators that involve Higgs and fermion fields. For instance, we have the following first-class operators:
\be
\widetilde{\cal O}^F_L = (i \bar F_L \lra\Dslash F_L)|H|^2 \ ,\quad
\widetilde{\cal O}_L^{(3)F} = (i \bar F_L \sigma^a\lra\Dslash F_L)(H^\dagger \sigma^a H)\ ,\quad
\widetilde{\cal O}^f_R =(i \bar f_R \lra\Dslash f_R)|H|^2\ ,
\label{Hfred}
\ee
as well as the second-class operators
\bea
&{\cal O}^u_{yH}= y_u D_\mu(\bar Q_L u_R) D^\mu \widetilde H\ , \quad &
{\cal O}^u_{yR}= y_u D_\mu(\bar Q_L \widetilde H)\gamma^\mu\gamma^\nu D_\nu u_R\ , \nonumber\\
&{\cal O}^u_{yL}= y_u D_\mu\bar Q_L\gamma^\mu\gamma^\nu D_\nu( \widetilde H u_R)\ ,
\quad &
{\cal O}^u_{yLR}=y_u (D_\mu \bar Q_L)\gamma^\mu\gamma^\nu (D_\nu u_R)\widetilde H\ ,
\label{Hfred2}
\eea
(and similar operators for down-type quarks and leptons). In addition, there are 
(second-class) operators involving fermions and gauge bosons:
\bea
&&{\cal O}_{BL}^F=g' (\bar F_L \gamma^\mu F_L)\partial^\nu B_{\mu\nu}\ ,\quad 
{\cal O}_{BR}^f=g' (\bar f_R \gamma^\mu f_R)\partial^\nu B_{\mu\nu}\ , \quad 
{\cal O}_{WL}^F=g (\bar F_L \sigma^a\gamma^\mu F_L)D^\nu W^a_{\mu\nu}\ ,\nonumber\\
&&
{\cal O}_{GL}^q=g_s (\bar Q_{L} T^A\gamma^\mu Q_{L})D^\nu G^A_{\mu\nu}\ ,\quad
{\cal O}_{GR}^f=g_s (\bar f_{R} T^A\gamma^\mu f_{R})D^\nu G^A_{\mu\nu}\ .
\label{Fgred}
\eea
To see that the operators (\ref{Hfred})-(\ref{Fgred})  can indeed be removed from the Lagrangian, consider the following field redefinitions that involve fermions:
\bea
Q_L&\rightarrow & Q_L (1+g_*^2\alpha_L J_H/\Lambda^2)+[g_*^2\alpha_L^{(3)} J^a_H\sigma^a Q_L
+i  \alpha^u_{yL}\Dslash J_{y_u\bar Q_L}+ y_u \alpha^u_{yLR}\widetilde H J_{Du_R}\nonumber\\
&&
+i \alpha^d_{yL}\Dslash J_{y_d\bar Q_L}+y_d \alpha^d_{yLR} H J_{D d_R}]/\Lambda^2\ ,
\nonumber\\
u_R & \rightarrow & u_R (1+g_*^2\alpha^u_R J_H/\Lambda^2)
+i \alpha^u_{yR}\Dslash J_{y_u \bar u_R}/\Lambda^2\ ,
\nonumber\\
d_R & \rightarrow & d_R (1+g_*^2\alpha^d_R J_H/\Lambda^2)
+i  \alpha^d_{yR}\Dslash J_{y_d \bar d_R}/\Lambda^2\ ,
\nonumber\\
B_\mu &\rightarrow &
B_\mu +g'[\alpha^B_{F_L} J_{F\mu}+\alpha^B_{f_R} J_{ff\mu}
]/\Lambda^2\ ,\nonumber\\
W^a_\mu &\rightarrow & 
W^a_\mu +g \alpha^W_{f_L} J^a_{L\mu} /\Lambda^2\ ,\nonumber\\
G^A_\mu &\rightarrow & 
G^A_\mu +g_s [\alpha^G_{F_{L}} J^A_{F\mu}+
\alpha^G_{f_{R}} J^A_{f\mu}] /\Lambda^2\ ,\nonumber\\
\widetilde H&\rightarrow & \widetilde H +\alpha_{Ht} J_{y_t H}^\dagger/\Lambda^2\ ,\nonumber\\
H&\rightarrow & H + \alpha_{Hb}J_{y_b H}^\dagger/\Lambda^2
\ ,
\eea
under which the Wilson coefficients shift as follows:
For the Higgs-fermion operators of Eq.~(\ref{first6dimF}), plus the straightforward generalization to the down-type fermions, and the up-down mixed operator of Eq.~(\ref{Hud}), we get (for third-generation quarks):
\bea
c_{y_t} &\rightarrow &  c_{y_t} -\alpha_1 -2\frac{\lambda}{g_*^2} \alpha_{Ht}-
\alpha^t_R-\alpha_L+\alpha_L^{(3)}\ ,
\nonumber\\
c_{y_b} &\rightarrow &  c_{y_b} -\alpha_1 -2\frac{\lambda}{g_*^2} \alpha_{Hb}-
\alpha^b_R-\alpha_L-\alpha_L^{(3)}\ ,
\nonumber\\
c_R &\rightarrow & c_R +\frac{|y_t|^2}{g_*^2} \alpha^t_{yL} + \frac{{g'}^2}{2g_*^2} (2Y_{R}^t \alpha_B+\alpha^B_{t_R})\ ,
\nonumber\\
c^{q_3}_L &\rightarrow & c^{q_3}_L -\frac{1}{2}\frac{|y_t|^2}{g_*^2} \alpha^t_{yR} + \frac{{g'}^2}{2 g_*^2} (2Y_{L}^q \alpha_B+\alpha^B_{Q_L})\ ,
\nonumber\\
c_L^{(3)q_3} &\rightarrow & c_L^{(3)q_3} +\frac{1}{2}\frac{|y_t|^2}{g_*^2} \alpha^t_{yR} + \frac{g^2}{2g_*^2} (\alpha_W+\alpha^W_{Q_L})\ , \nonumber\\
c^{tb}_R&\rightarrow & c_R^{tb}-\frac{1}{2}(\alpha^t_{yL}+\alpha^b_{yL})\ .
\eea
The Higgs-fermion redundant operators of Eq.~(\ref{Hfred}) can be eliminated by the shifts:
\bea
\tilde c^t_R&\rightarrow & \tilde c^t_R+\alpha^t_R -\frac{|y_t|^2}{g_*^2} (\alpha^t_{yL}+\alpha^t_{yLR})\ , 
\nonumber\\
\tilde c^b_R&\rightarrow & \tilde c^b_R+\alpha^b_R \ , 
\nonumber\\
\tilde c^{q_3}_L&\rightarrow & \tilde c^{q_3}_L+\alpha_L -\frac{1}{2}\frac{|y_t|^2}{g_*^2} \alpha^t_{yR}\ , 
\nonumber\\
\tilde c^{(3)q_3}_L&\rightarrow & \tilde c^{(3)q_3}_L+\alpha^{(3)}_L +\frac{1}{2}\frac{|y_t|^2}{g_*^2} \alpha^t_{yR}\ ,
\eea
(where, from here on, we neglect $|y_b|^2$ and $|y_\tau|^2$ contributions) while the redundant Higgs-fermion operators of Eq.~(\ref{Hfred2}) can be eliminated by the shifts:
\bea
c^f_{yH}&\rightarrow & c^f_{yH}+\alpha_{Hf}+ \alpha_2\ ,\nonumber\\
c^f_{yR}&\rightarrow & c^f_{yR}+\alpha^f_{yR}\ ,\nonumber\\
c^f_{yL}&\rightarrow & c^f_{yL}+\alpha^f_{yL}\ ,\nonumber\\
c^f_{yLR}&\rightarrow & c^f_{yLR}+\alpha^f_{yLR}\ .
\eea
All the redundant gauge-fermion operators of Eq.~(\ref{Fgred}) can be removed by the shifts:
\bea
c^f_{BR}&\rightarrow & c^f_{BR}+Y_{R}^f\alpha_{2B}- \alpha^B_{f_R}\ ,\nonumber\\
c^F_{BL}&\rightarrow & c^F_{BL}+Y_{L}^F\alpha_{2B}- \alpha^B_{F_L}\ ,\nonumber\\
c^F_{WL}&\rightarrow & c^F_{WL}+\frac{1}{2}\alpha_{2W}- \alpha^W_{F_L}\ ,\nonumber\\
c^q_{GL,GR}&\rightarrow & c^q_{GL,GR}+\alpha_{2G}-\alpha^G_{q_{L,R}}\ .
\eea
Finally, the coefficients of four-fermion operators will also be shifted but we will not need such shifts and we do not list them.

Using all the shift freedom to remove these redundant operators we end up (say, for the third family) with the following Higgs-fermion Wilson coefficients: $y_fc_{y_f}, c^f_R, c^F_L, c_L^{(3)F}$ and $c_R^{tb}$, with $f=t,b,\tau$ and $F=q,l$, in agreement with the operators listed in Table 2 of Section~\ref{basis}.

\newpage
\section{Anomalous dimensions of $\bma{d=6}$ Wilson coefficients} 
\label{AppRGEs}

In our analysis we are interested in potentially large radiative effects in the running of the $d=6$ Wilson coefficients $c_i$ from the scale $\Lambda$ of new physics to the electroweak scale. To study such effects we have computed the one-loop anomalous dimensions $\gamma_{c_i}$ for the Wilson coefficients, which are functions of the coefficients themselves, that is:
\be
\gamma_{c_i}=\frac{d c_i}{d \log \mu} = \gamma_{c_i}(c_j)\ ,
\label{gammas}
\ee
where $\mu$ is the renormalization scale.

When redundant operators are removed from the Lagrangian some care has to be taken in computing anomalous dimensions of the operators left in the basis. The reason is that redundant operators
can be generated through RG evolution by operator mixing with 
non-redundant operator. In other words, the $\gamma_{c_i}$'s of redundant operators are not zero in general. 

Let us explain how this effect can be taken care of in a simple way.
Consider a basis formed by a set of coefficients $\{c_i\}$, after removing a set of redundant coefficients $\{c^r_i\}$. The procedure
to remove the $c^r_i$ is straightforward and 
has been illustrated in the previous Appendix.
One starts from the shifts induced by field-redefinitions with arbitrary parameters $\alpha_k$, which have the form
\be
c_i \rightarrow c'_i (\alpha_j)= c_i +\sum_k s_{ik} \alpha_k\ ,\quad
c^r_i \rightarrow  {c^r}'_i (\alpha_j)= c^r_i +\sum_k s^r_{ik} \alpha_k\ .
\label{cishifts}
\ee 
Then the $\alpha_k$'s are chosen so as to remove the redundant operators,
\be
{c^r}'_i (\alpha^*_j)=0\;\;\; \Rightarrow \;\;\; \alpha^*_j = -\sum_i (s^r)^{-1}_{ji}
c^r_i\ .
\ee
It is then convenient to define the following combinations 
\be
C_i \equiv c_i'(\alpha^*_j) = c_i - \sum_{kl} s_{ik}(s^r)^{-1}_{kl}c^r_l\ ,
\ee
which are invariant under the arbitrary shifts (\ref{cishifts}) and correspond to a more physical definition of the Wilson coefficients.

The anomalous dimensions of these shift-invariant $C_i$'s are simply
\be
\gamma_{C_i} = \gamma_{c_i} - \sum_{kl} s_{ik}(s^r)^{-1}_{kl}\gamma_{c^r_l}=\gamma_i(c_k;c_k^r)\ ,
\label{gammaCi}
\ee
where the last expression just indicates some function of the Wilson coefficients and we distinguish in its argument between coefficients in the basis and coefficients of redundant operators.
The key property of this function is that it must depend on the Wilson coefficients only through the shift-invariant combinations. That is,
it satisfies
\be
\gamma_i(c_k;c^r_k) = \gamma_i(C_k;0)\ . 
\ee
This implies that setting $c_k^r=0$ in these $\gamma_i(c_k;c^r_k)$
functions is now a consistent procedure to obtain the anomalous dimensions after removing redundant operators.\footnote{This is equivalent to the procedure described independently in \cite{EinhornWudka}.} An explicit example of this is given at the end of Appendix~\ref{AppRGEs}. We applied this procedure to calculate  the anomalous dimensions used in Section~\ref{sec:RGEs}. In the next Subsections we will list the
required shift-invariant $C_i$ combinations and  present the $\gamma_{c_i}$'s necessary to complete the
calculation.

\subsection{Shift-invariant combinations of Wilson coefficients}

In order to simplify the expressions for the shift-invariant 
combinations $C_i$ of Wilson coefficients, we present them first in a basis that treats as redundant the operators ${\cal O}_B$,  ${\cal O}_W$, ${\cal O}_{2B}$, ${\cal O}_{2W}$ and  ${\cal O}_{2G}$.
We explain afterwards how to express these combinations in other bases, as  those that keep these operators. As we are not interested in this paper in
calculating the anomalous dimensions of 4-fermion operators, below we restrict our $C_i$'s to non-4-fermion current-current operators.
We find:
\bea
C_H&\equiv & c_H - c_r -\frac{3g^2}{4g_*^2}(2c_W-c_{2W})\ ,
\nonumber\\
C_T&\equiv & c_T-\frac{{g'}^2}{4 g_*^2}(2c_B-c_{2B})\ ,\nonumber\\
C_6&\equiv & c_6 +2c_r+\frac{g^2}{g_*^2}(2c_W-c_{2W})+4 \frac{\lambda}{g_*^2}c_{K4}\ ,\nonumber\\
C_{y_t}&=&c_{y_t}+\frac{1}{2}c_r+2\frac{\lambda}{g_*^2}\left(c^t_{yH}+ c_{K4}\right)+\tilde c^t_R+\tilde c^{q_3}_L-\tilde c_L^{(3)q_3}
+\frac{g^2}{4g_*^2}(2c_W-c_{2W})
\nonumber\\
&+&\frac{|y_t|^2}{g_*^2}(c^t_{yR}+c^t_{yL}+c^t_{yLR})\ ,
\nonumber\\
C_{y_b}&=& c_{y_b}+\frac{1}{2}c_r+2\frac{\lambda}{g_*^2}\left(c^b_{yH}+ c_{K4}\right)+\tilde c^b_R+\tilde c^{q_3}_L+\tilde c_L^{(3)q_3}
+\frac{g^2}{4g_*^2}(2c_W-c_{2W})\ ,\nonumber\\
C_{y_\tau}&=& c_{y_\tau}+\frac{1}{2}c_r+2\frac{\lambda}{g_*^2}\left(c^\tau_{yH}+ c_{K4}\right)+\tilde c^\tau_R+\tilde c^{l_3}_L+\tilde c_L^{(3)l_3}
+\frac{g^2}{4g_*^2}(2c_W-c_{2W})\ ,\nonumber\\
C^t_R&=&c^t_R-\frac{|y_t|^2}{g_*^2}c^t_{yL}+\frac{{g'}^2}{2g_*^2}[Y_{R}^t(c_B-c_{2B})+c^t_{BR}]\ ,
\nonumber\\
C_R^b&=&c^b_R+\frac{{g'}^2}{2g_*^2}[Y_{R}^b(c_B-c_{2B})+c^b_{BR}]\ ,
\nonumber\\
C_R^\tau&=&c^\tau_R+\frac{{g'}^2}{2g_*^2}[Y_{R}^\tau(c_B-c_{2B})+c^\tau_{BR}]\ ,
\nonumber\\
C^{q_3}_L&=&c^{q_3}_L+\frac{1}{2}\frac{|y_t|^2}{g_*^2} c^t_{yR}+\frac{{g'}^2}{2g_*^2}[Y_{L}^q(c_B-c_{2B})+c^{q_3}_{BL}]\ ,
\nonumber\\
C^{l_3}_L&=&c^{l_3}_L+\frac{{g'}^2}{2g_*^2}[Y_{L}^l(c_B-c_{2B})+c^{l_3}_{BL}]\ ,
\nonumber
\eea
\bea
C_L^{(3)\, q_3}&=&c_L^{(3)q_3}-\frac{1}{2}\frac{|y_t|^2}{g_*^2}c^t_{yR}+\frac{g^2}{4 g_*^2}(c_W-c_{2W}+2c^{q_3}_{WL})\ ,\nonumber\\
C_L^{(3)\, l_3}&=&c_L^{(3)l_3}+\frac{g^2}{4 g_*^2}(c_W-c_{2W}+2c^{l_3}_{WL})\ ,\nonumber\\
C_R^{tb}&=&c_R^{tb}+\frac{1}{2}(c_{yL}^t+c_{yL}^b)\ .
\label{physcoef}
\eea
Out of the 59 independent operators for a single family, 20 are of one-loop type and  25 are 4-fermion tree-level operators. The remaining 14 are  tree-level operators whose number corresponds to the 14 physical 
$C_i$'s in (\ref{physcoef}). 

Let us now discuss how these $C_i$'s would be modified in other basis.
For example, if we keep ${\cal O}_B$ and ${\cal O}_W$ in the basis instead of the leptonic operators ${\cal O}_L^l$ and ${\cal O}_L^{(3)\, l}$, then one should remove $C_L^l$ and $C_L^{(3)l}$ from the list of $C_i$'s.
This is accomplished by making the replacements 
\bea
c_B &\rightarrow & c_{2B} -\frac{1}{Y_L^l} c_{BL}^{l}-\frac{2g_*^2}{Y_L^l {g'}^2}c_L^l\ ,
\nonumber\\
c_W &\rightarrow & c_{2W} - 2 c_{WL}^l - 4\frac{g_*^2}{g^2}c_L^{(3)\, l}\ ,
\label{Cichange}
\eea
in all the $C_i$'s (obtaining in particular $C_L^l=C_L^{(3)\, l}=0$) and then add to the list the following two new $C_i$'s:
\bea
C_B&=& c_B - c_{2B} +\frac{1}{Y_L^l} c_{BL}^l+\frac{2g_*^2}{Y_L^l {g'}^2}c_L^l\ ,
\nonumber\\
C_W&=& c_W - c_{2W} + 2 c_{WL}^l + 4\frac{g_*^2}{g^2}c_L^{(3)\, l}\ .
\eea
The replacement in \eq{Cichange} introduces a dependence on $\gamma_{c_{BL}^l}$, $\gamma_{c_{c_L}^l}$, $\gamma_{c_{WL}^l}$ and $\gamma_{c_L^{(3) l}}$ in the calculation of the anomalous dimensions of the $C_i$'s, but the only non-redundant coefficients that appear in those anomalous dimensions depend on leptonic Yukawa couplings that we neglect.

In a similar way, ${\cal O}_{2B}$, ${\cal O}_{2W}$ and ${\cal O}_{2G}$ can be kept in the basis instead of three 4-fermion operators of the first family, {\em e.g.} ${\cal O}_{RR}^e$, ${\cal O}_{LL}^l$ and ${\cal O}_{RR}^{(8)d}$.  In this basis $c_{2B}, c_{2W}$ and $c_{2G}$ have to be replaced by linear combinations of $c_{RR}^e, c_{LL}^l$ and $c_{RR}^{(8)d}$ in the $C_i$'s above. However, this replacement has no impact on the anomalous dimensions of the $C_i$'s if we only keep the coefficients of \eq{cjs} and neglect small Yukawas. Indeed, it is simple to realize that  \eq{cjs}  can only renormalize $c_{2B,2W,2G}$ or  $c_{RR}^e, c_{LL}^l$ and $c_{RR}^{(8)d}$
through lepton or down Yukawas, which are terms we neglect in our RGEs.
Therefore, whether we keep ${\cal O}_{2B,2W,2G}$ or 4-fermion operators, the RGEs given in the main body of the paper are unaffected.

\subsection{Anomalous dimensions before removing redundant operators}

To calculate the anomalous dimensions $\gamma_{C_i}$'s, following \eq{gammaCi}, we need 
to calculate the anomalous dimensions of the Wilson coefficients entering in the $C_i$'s, including those that are redundant.

We have calculated these anomalous dimensions to linear order in the $c_j$'s of \eq{cjs}, the only exception being $c_r$, which we keep for illustrative purposes here. Parametrically one has $\gamma_{c_i} \sim g_j^2 c_j/16\pi^2$ and we only keep $g_j^2 =\{y_t^2, g_s^2, g^2, {g'}^2,\lambda\}$, dropping $g_j^2=\{y_b^2, y_\tau^2,...\}$. The anomalous dimensions, calculated in Landau gauge,  are:
\bea
\gamma_{c_H}&=&\frac{1}{4\pi^2}\left\{N_c y_t^2 [c_{y_t}+c_L^{(3)}]+\lambda(7c_H-c_r)
+\frac{3}{8}\left[g^2 (c_H+2c_r)+{g'}^2c_r\right]
\right\}-4\gamma_h c_H , \label{B10}
\\[0.2cm]
\gamma_{c_T}&=& \frac{1}{16\pi^2}\left[
4N_c y_t^2(c_R-c_L)+\frac{3}{2}{g'}^2(c_H-c_r) \right]-4 \gamma_h c_T\ , 
\\[0.2cm]
\gamma_{\lambda c_6}&=&\frac{1}{8\pi^2}\left\{54\lambda^2c_6 -4N_cy_t^4c_{y_t}+12\lambda^2(3c_H+2c_r)-\frac{3}{8}\left[2g^4+(g^2+{g'}^2)^2\right]c_r\right\}-6\gamma_h \lambda c_6\ ,\nonumber\\
&&\\
\frac{1}{y_t}\gamma_{y_tc_{y_t}}&=&
\frac{1}{16\pi^2}\left\{4\lambda\left[
c_R-c_L+3c_L^{(3)}+6c_{y_t}\right]-{g'}^2\left[c_R+4c_L-4c_L^{(3)}+\frac{2}{3}c_{y_t}\right]
\right.\nonumber\\
&&\left.-3g^2c_R-8g_s^2c_{y_t}
+2y_t^2\left[4c_{LR}+4C_F c_{LR}^{(8)}+c_R-c_L+ c_L^{(3)}+2c_{y_t}
+c_H\right]
\right\}\nonumber\\
&&
-(3\gamma_h+\gamma_{Q_L}
+\gamma_{t_R})c_{y_t}\ ,
\\[0.2cm]
\frac{1}{y_b}\gamma_{y_b c_{y_b}} & = &\frac{1}{8\pi^2}\left\{ 2 \lambda [c_L+3 c_L^{(3)}+6c_{y_b} ] +y_t^2[2 c_L^{(3)}-c_{y_t}]
+\left(\frac{1}{6} g^{\prime 2}  - 4 g_s^2 \right) c_{y_b}   \right.
\nonumber \\
& &\left.  +\frac{y_t^2}{g_*^2}[3g^2- 2y_t^2-4\lambda]c^{tb}_{R} +g^{\prime 2} [c_L+c_L^{(3)}]
-\frac{y_t^4}{g_*^2}\left[ \left(2N_c + 1\right)   c_{y_t y_b}+C_F c^{(8)}_{y_t y_b} \right] \right\}
\nonumber\\
 &&-  (\gamma_{Q_L}+\gamma_{b_R}+3 \gamma_h) c_{y_b} \ ,
\\[0.2cm]
  \frac{1}{y_\tau} \gamma_{ y_\tau c_{y_\tau}}  & =& \frac{1}{16\pi^2}\left[
3\left(8 \lambda - g^{\prime 2}     \right) c_{y_\tau}
+2N_c\frac{y_t^2}{g_*^2} (\lambda-y_t^2)(2c_{y_ty_\tau}+
c'_{y_ty_\tau})
\right]\nonumber\\
&&-  (\gamma_{L_L}+\gamma_{\tau_R}+3 \gamma_h)  c_{y_\tau}  \ ,
  \\[0.2cm]
\gamma_{c_R}&=&\frac{1}{8\pi^2}\left\{
y_t^2\left[N_c c_{LR}-2(N_c+1)c_{RR}+2c_R-c_L
+\frac{1}{4}(c_H-c_r)\right]-\frac{3}{4}(3g^2+{g'}^2)c_R\right\}\nonumber\\
&&-2(\gamma_h+\gamma_{t_R})c_R\ ,
\\[0.2cm]
\gamma_{c_L}&=&\frac{1}{8\pi^2}\left\{y_t^2\left[-N_c c_{LR}
-\frac{1}{2}c_R +c_L-3c_L ^{(3)}-\frac{1}{8}(c_H-c_r)+(2N_c+1)c_{LL}+C_F\, c_{LL}^{(8)}\right]
\right.\nonumber\\
&&\left.-\frac{3}{4}(3g^2+{g'}^2)c_L
\right\}-2(\gamma_h+\gamma_{Q_L})c_L\ ,
\\[0.2cm]
\gamma_{c_L^{(3)}}&=&\frac{-1}{8\pi^2}
\left\{y_t^2\left[c_L+c_L ^{(3)}-\frac{1}{8}(c_H-c_r)
+c_{LL}+C_F\, c_{LL}^{(8)}\right]+\frac{3}{4}(g^2+{g'}^2) c_L^{(3)}\right\}\nonumber\\
&&-2(\gamma_h+\gamma_{Q_L})c_L^{(3)}\\[0.2cm]
\gamma_{c_r}&=&\frac{1}{4\pi^2}\left\{N_c y_t^2 \left[c_{y_t}-2c_L^{(3)}\right]+\lambda(c_H+5c_r)+\frac{3}{8}\left[(5g^2+{g'}^2)c_r-2g^2 c_H\right]
\right\}-4\gamma_h c_r\ ,\nonumber
\eea
\bea
\frac{1}{y_t}\gamma_{y_t c_{yH}}&=&-\frac{g_*^2}{4\pi^2}\left[ c_{LR}+C_F\, c_{LR}^{(8)}
\right]-(\gamma_h+\gamma_{Q_L}+\gamma_{t_R})
c_{yH}\ ,\\[0.2cm]
\frac{1}{y_t}\gamma_{y_t c_{yR}}&=&\frac{g_*^2}{8\pi^2}\left[c_L-3c_L^{(3)}\right]-(\gamma_h+\gamma_{Q_L}+\gamma_{t_R})
c_{yR}\ ,\\[0.2cm]
\frac{1}{y_t}\gamma_{y_t c_{yL}}&=&-\frac{g_*^2}{8\pi^2}c_R-(\gamma_h+\gamma_{Q_L}+\gamma_{t_R})
 c_{yL}\ ,\\[0.2cm]
\frac{1}{y_t}\gamma_{y_t c_{yLR}}&=&\frac{g_*^2}{16\pi^2}\left[c_R-c_L+3c_L^{(3)}\right]-(\gamma_h+\gamma_{Q_L}+\gamma_{t_R})
c_{yLR} \ , 
\\[0.2cm]
\frac{1}{y_b}\gamma_{y_b c^b_{yH}}&=&\frac{1}{16\pi^2}y_t^2\left[ \left(2N_c+1\right)c_{y_ty_b}+C_F\, c_{y_ty_b}^{(8)}
\right]-(\gamma_h+\gamma_{Q_L}+\gamma_{b_R})
c^b_{yH}\ ,\\[0.2cm]
\gamma_{c_{K4}}&=&\frac{y_t^2}{4\pi^2}N_c c_{yH}-2\gamma_h c_{K4}\ ,
\\[0.2cm]
\gamma_{\tilde c_R}&=&\frac{y_t^2}{16\pi^2}\left[3c_{y_t}+c_R+\frac{1}{2}(c_H+2c_r)\right]-2(\gamma_h+\gamma_{t_R})\tilde c_R\ ,
\\[0.2cm]
\gamma_{\tilde c_L}&=&\frac{y_t^2}{32\pi^2}\left[3c_{y_t}-c_L+3c_L^{(3)}+\frac{1}{2}(c_H+2c_r)\right]-2(\gamma_h+\gamma_{Q_L})\tilde c_L\  ,
\\[0.2cm]
\gamma_{\tilde c_L^{(3)}}&=&\frac{y_t^2}{32\pi^2}\left[-c_{y_t}+c_L+c_L^{(3)}-
\frac{1}{2}c_H\right]-2(\gamma_h+\gamma_{Q_L})\tilde c_L^{(3)}\  ,
\\[0.2cm]
\gamma_{c_W}&=&\frac{g_*^2}{48\pi^2}\left[
16N_c c_L^{(3)}-(c_H+c_T)\right] -\left(2\gamma_h + \gamma_W + \frac{1}{g}\beta_g\right)c_W , 
\\[0.2cm]
\gamma_{c_B}&=& \frac{g_*^2}{48\pi^2}\left[\frac{8N_c}{3}(2c_R+c_L) -
(c_H+5c_T)\right]-2\gamma_h c_B , 
\\[0.2cm]
\gamma_{c_{BR}}&=& \frac{g_*^2}{12\pi^2}\left\{\frac{1}{3}\left[4(N_c+1)c_{RR}+N_c c_{LR}\right]+c_R\right\}-2\gamma_{t_R}c_{BR},
\\[0.2cm]
\gamma_{c_{BL}}&=& \frac{g_*^2}{12\pi^2}\left\{\frac{1}{3}\left[(2N_c+1)c_{LL}+C_Fc_{LL}^{(8)}+N_c c_{LR}\right]+2c_L\right\}-2\gamma_{Q_L}c_{BL},
\\[0.2cm]
\gamma_{c_{WL}}&=& \frac{g_*^2}{12\pi^2}\left[c_{LL}+C_F c_{LL}^{(8)}+c_L^{(3)}\right]-\left(2\gamma_{Q_L}+ \gamma_W + \frac{1}{g}\beta_g\right)c_{WL}, \label{B33}
\eea
where $C_F=(N_c^2-1)/(2N_c)$, $\beta_g=\frac{d g}{d \log \mu}$  and 
\bea
&&
\gamma_h =\frac{1}{16\pi^2}\left[-N_c y_t^2+\frac{3}{4}(3g^2+{g'}^2)\right],\quad  
\gamma_{Q_L}=\frac{1}{16\pi^2}\left[-\frac{1}{2}y_t^2\right] \ ,
\quad \gamma_{t_R}=- \frac{y_t^2}{16\pi^2}\ ,\nonumber\\
&&\gamma_W = - \frac{1}{g}\beta_g - \frac{3}{16\pi^2} g^2 = \frac{1}{16\pi^2} \frac{g^2}{6} , 
\eea
are the wave-function renormalization terms. The corresponding wave-function terms for leptons and $b_R$ ($\gamma_{L_L}, \gamma_{\tau_R}$ and $\gamma_{b_R}$) are proportional to small Yukawa couplings squared that we are neglecting.
 Notice that in the above results we have included some dependence on Wilson coefficients beyond those of  \eq{cjs} and $c_r$. In particular, we have kept the contributions from wave function renormalization (which are trivial to take into account)  in all cases, and 
we also kept the contributions from $c_T$ in $\gamma_{c_W}$ and $\gamma_{c_B}$ that were already calculated in \cite{us}.
These anomalous dimensions have been calculated through the (divergent pieces of the) one-loop effective action.

\begin{figure}[t]
\begin{center}
\includegraphics[width=0.85\textwidth]{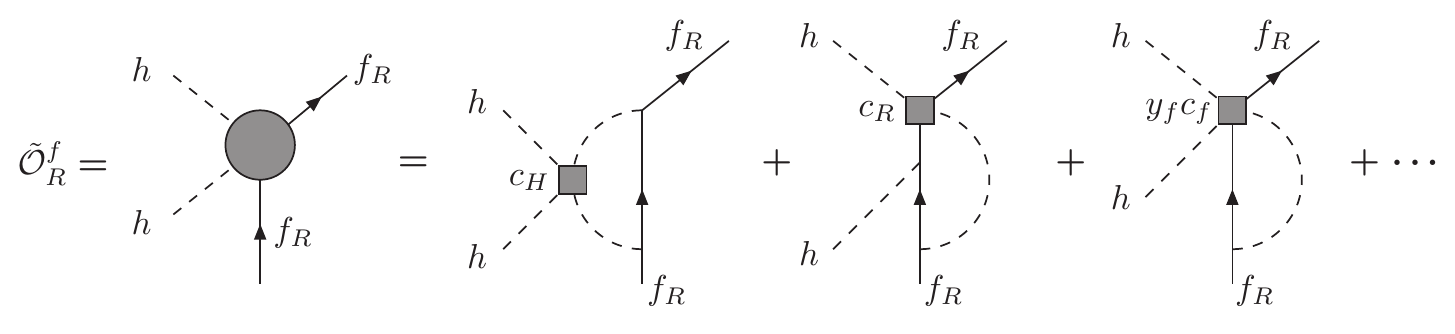}   
\parbox{14cm}{
\caption{\label{redundiags1}
\emph{Diagrams that generate at one-loop the redundant operator $\tilde {\cal O}_R^f$. }}}
 \end{center}
\end{figure}

Using (\ref{B10})-(\ref{B33}) we can calculate the anomalous dimensions $\gamma_{C_i}$'s for the shift-invariant Wilson coefficients. These are given in Section~\ref{sec:RGEs}. 
We have cross-checked those RGEs by calculating them in an alternative way. We have looked at the one-loop radiative corrections to some particular physical processes and required the corresponding amplitudes to be independent of the renormalization scale. In order to find agreement between both methods, it is crucial to include in the amplitude for the physical process non-1PI contributions. In the effective action approach, such diagrams are in one-to-one correspondence with the redundant operators being eliminated.

\begin{figure}[t]
\begin{center}
\includegraphics[width=0.8\textwidth]{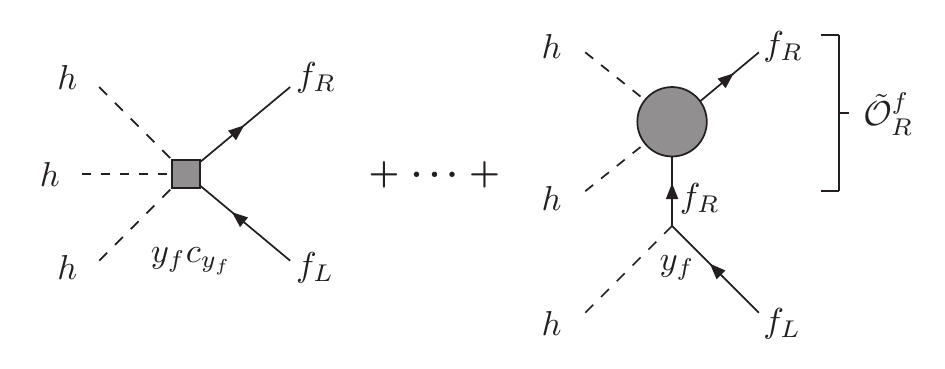}
\parbox{14cm}{
\caption{\label{redundiags2}
\emph{Contributions to the process $hhh\rightarrow \bar f_L f_R$ at order $1/\Lambda^2$ , including the one-loop corrections shown in Fig.~\ref{redundiags1}.}}}
 \end{center}
\end{figure}
As an illustrative example of the previous point, consider the contribution of the redundant operator $\tilde{\cal O}_R^f$ to the renormalization of ${\cal O}_{y_f}$. 
One-loop radiative corrections do generate $\tilde{\cal O}_R^f$ in the one-loop effective action, as shown 
in Fig.~\ref{redundiags1}, even if we remove $\tilde{\cal O}_R^f$ from the (tree-level) Lagrangian. 
The physical combination $C_{y_f}$ [see Eq.~(\ref{physcoef})] depends on ${\tilde c}_R^f$  and, therefore, the anomalous dimension
$\gamma_{C_{y_f}}$ also depends on $\gamma_{\tilde c_R^f}$. 
The same result for $\gamma_{C_{y_f}}$ can be obtained by looking at the physical process $hhh\rightarrow \bar f_L f_R$. The $1/\Lambda^2$ diagrammatic contributions to this process are shown in Fig.~\ref{redundiags2}. Besides the tree-level contribution through $c_{y_f}$ shown on the left, there are one-loop corrections, among which we just show the ones related to the redundant operator $\tilde{\cal O}_R^f$. Having removed the redundant 
$\tilde{\cal O}_R^f$ from the basis, there is no tree-level contribution from 
$\tilde c_R^f$  to $hhh\rightarrow \bar f_L f_R$ and the divergences from the one-loop blob shown in Fig.~\ref{redundiags2} have to be absorbed by $c_{y_f}(\mu)$ to obtain an amplitude that is independent of the renormalization scale $\mu$. 

Finally, the reader can check, using the previous anomalous dimensions which include the dependence on the redundant coefficient $c_r$, that the anomalous dimensions of the shift invariant combinations $c_H-c_r$, $c_6+2c_r$,
$c_{y_t}+c_r/2$, $c_{y_b}+c_r/2$, plus all the other Wilson coefficients, are functions of these same combinations, so that one can take $c_r=0$ in a consistent way.

\newpage
\section{Transformation properties of the $\bma{d=6}$ operators\\
under the custodial ${\bma{SU(2)_L}}{\bma{\otimes}} {\bma{SU(2)_R}}$ symmetry} 
\label{appc}

The $d=6$ operators of the basis of Section~2 can be made invariant under the custodial ${SU(2)_L}\otimes {SU(2)_R}$ by promoting their coefficients to (non-propagating) spurion   fields transforming  
under this symmetry.  In this Appendix we present these transformation rules.

The  bosonic sector of the SM Lagrangian can be made custodial invariant by  promoting
the  gauge coupling $g'$  to transform as  a  triplet  under $SU(2)_R$:
\be
g^{\prime}_a \sigma_a \rightarrow  g^{\prime }_a\, R  \sigma_a R^\dagger\, ,
\label{spurion}
 \ee
whose  nonzero  VEVs,  given by $\langle g^{\prime }_a\rangle=g^{\prime} \delta_{a,3} $,
define how the custodial symmetry is explicitly broken by this coupling.
For the Higgs field, that  transforms as a  $({\bf 2}_L,{\bf 2}_R)$, 
it is convenient to use  the  matrix field
 \be
 \Sigma = \frac{1}{\sqrt{2}} \left( \tilde H\, ,\, H\right)\, ,
 \ee
that  transforms under  the custodial group as  $\Sigma \rightarrow L \Sigma R^\dagger$, and  therefore its covariant derivative is 
given by
$D_\mu\Sigma=\partial_\mu\Sigma-igW_{L\, \mu}^a\sigma^a\Sigma/2+ig^{\prime\, a}B_\mu \Sigma  \sigma^a/2$.

To write  the Yukawa sector of the SM invariant under the custodial symmetry, we can promote the Yukawa couplings
to transform as a doublet under $SU(2)_R$:
\be 
  Y_u \rightarrow R Y_u\, ,
\label{spurion}
 \ee
where $\langle Y_u\rangle =(y_u, 0)^T$, and similarly for the other Yukawas.
The  Yukawa term is then written as $ \sqrt{2}  \bar Q_L \Sigma Y_u u_R$, where  the SM fermions transform as  singlets under $SU(2)_R$.
To define the proper hypercharge assignment for the SM fermions, we have to enlarge the global group 
to  ${SU(2)_L}\otimes {SU(2)_R}\otimes U(1)_X$ and define the hypercharge as $Y=T_R^3+X$.
This means that   the $U(1)_Y$ is not only contained in $SU(2)_R$ but also in $U(1)_X$ and therefore $g'$ has also a singlet component under the custodial group.
 
Using the above definitions we can  write the Lagrangian ${\cal L}_6$  as an invariant  under ${SU(2)_L}\otimes {SU(2)_R}$. This requires to  promote few of the coefficients  to spurion fields transforming  non-trivially. The result is   shown in Table~\ref{custoinv}. Only $c_T$ and $c_{L,R}$ transform non-trivially, being then, together with  $g'$ and the Yukawa couplings, the only  sources of 
custodial breaking. This information is useful to deduce what combinations of coefficients and couplings can contribute at the one-loop level to a given anomalous dimension.
For example, contributions to  $\gamma_{c_T}$ can only come from terms that transform as    $\bm{5}_R$:  $\langle g^{\prime}_a g^{\prime}_b c_H\rangle=g^{\prime\, 2}c_H\delta_{a,3}\delta_{b,3}$ and $\langle Y_u^\dagger \sigma^a Y_u c^b_{L,R}\rangle=-y_u^2 c_{L,R}\delta^{a,3}\delta^{b,3}$,
as the explicit calculation shows. In the same way it can be understood 
why $\gamma_{c_H}$ depends on $y_t^2c_L^{(3)}$ but not on $y_t^2c_L$,  being $c_H$ a singlet under the custodial symmetry.

Useful information can also be derived from the transformations under 
the parity $P_{LR}$ that interchanges $L\leftrightarrow R$.
In the bosonic sector, we have
\bea
\Sigma&\leftrightarrow& \Sigma^\dagger\nonumber
\eea
\bea
\frac{g^{\prime\, a}}{g'}B_\mu &\leftrightarrow& W_{L\, \mu}^a\nonumber\\
g'&\leftrightarrow& g\nonumber\\
c_H&\leftrightarrow& c_H\nonumber\\
c_W&\leftrightarrow& c_B\nonumber\\
\kappa_{HW}&\leftrightarrow& \kappa_{HB}\, ,
\label{prlb}
\eea
while $c_T$ and $\kappa_{BB}$ do not have a well-defined transformation property inside the operator basis.
For this reason it could be convenient to work with the operator ${\cal O}_{WB}$ instead of
 ${\cal O}_{BB}$  [both related by \eq{OpId2}] that is even under $P_{LR}$, and therefore  $\kappa_{WB}\leftrightarrow \kappa_{WB}$.

For  operators involving SM fermions, we have several possibilities for the transformation properties under
$P_{LR}$, see \cite{custodial}.
The two simplest ones are  to consider (for the up-type quark)
\be
{\rm  I)}\ Q_R\equiv\frac{1}{y_u} Y_u u_R\ \ \ {\rm and}\ \ \ \ Q_L
\label{firstchoice}
 \ee
  that transform respectively as  $(\bm 1,\bm 2_R)_{1/6}$ 
  and $(\bm 2_L,\bm 1)_{1/6}$
  under ${SU(2)_L}\otimes {SU(2)_R}\otimes U(1)_X$,
  or, alternatively, 
 \be
{\rm  II)}\ {\cal Q}_L\equiv\frac{1}{y_u} Q_L\otimes Y_u^\dagger\ \  \ {\rm and}\ \ \ u_R
\label{secondchoice}
 \ee
transforming  as   $(\bm 2_L,\bm 2_R)_{2/3}$ and   $(\bm 1,\bm 1)_{2/3}$ respectively.
For the first case, \eq{firstchoice}, 
we  can write the operators ${\cal O}_R$ and ${\cal O}_L^{(3)}$ in the following way: 
\be
-i c_R \text{tr}[  \sigma_a \Sigma^\dagger  {\lra{D}_\mu} \Sigma] \bar{Q}_R \sigma_a \gamma^\mu Q_R
\ \ {\rm and}\ \ i c_L^{(3)} \text{tr}[  \Sigma^\dagger  \sigma_a {\lra{D}_\mu} \Sigma] \bar{Q}_L \sigma_a \gamma^\mu Q_L\ , 
\ee
such that  under $P_{LR}$ we can define $Q_L\leftrightarrow Q_R$  
and 
\be
{\rm I)}\ c_R\leftrightarrow  c_L^{(3)}\, .
\ee
For the second case, \eq{secondchoice},  we can write the operators  ${\cal O}_L$ and ${\cal O}_L^{(3)}$ as
\be
i c_L \text{tr}[  \sigma_a  \Sigma^\dagger  {\lra{D}_\mu} \Sigma ]\text{tr}[  \bar{\cal{Q}}_L^T \epsilon^T \sigma_a \gamma^\mu  \epsilon {\cal{Q}}_L^T   ]
\ \ {\rm and}\ \ i  c_L^{(3)} \text{tr}[  \Sigma^\dagger  \sigma_a {\lra{D}_\mu} \Sigma]\text{tr}[ \bar{\cal{Q}}_L \sigma_a \gamma^\mu {\cal{Q}}_L] \ ,
\ee
 and define  ${\cal Q}_L\leftrightarrow\epsilon {\cal Q}_L^T \epsilon^T$ under $P_{LR}$
that gives the transformation rule
\be
{\rm II)}\ c_L\leftrightarrow  - c_L^{(3)}\, .
\label{PLPRtobeused}
\ee
In this latter case,  invariance under  $P_{LR}$ implies  $c_L+c_L^{(3)}=0$,
and therefore no corrections to the $Zb_L\bar b_L$ coupling.

{\renewcommand{\arraystretch}{1.7} \renewcommand{\tabcolsep}{0.29cm}
\begin{table}
\centering
\small
\begin{tabular}{|c  | c |  c | c |}

\hline \textbf{Operator} & \textbf{Spurion} & $\bm{SU(2)_L}\otimes \bm{SU(2)_R}$ & \textbf{VEV} \\  \hline \hline 
 
 $\frac{1}{2} c_T^{a,b}  \text{tr}[\sigma^a \Sigma^\dagger {\lra{D}_\mu} \Sigma] \text{tr}[\sigma^b \Sigma^\dagger {\lra{D^\mu}} \Sigma]$   & $c_T^{a,b}$  &$ (\bm{3}_R\otimes \bm3_R)_{s}=\bm{5}_R+\bm 1$ & $c_T \delta^{a,3}\delta^{b,3} $\\ \hline

 $\frac{1}{2} c_H \left( \partial_\mu \text{tr}[ \Sigma^\dagger  \Sigma] \right)^2$   & $c_H$ & $\bm1$ & $c_H$ \\ \hline

  $c_6 \left(\text{tr}[\Sigma^\dagger \Sigma]\right)^3$   & $c_6$ & $\bm 1$ & $c_6$ \\ \hline

  $ -\frac{  i }{ 2} c_B g^{\prime a} \text{tr}[ \sigma^a  \Sigma^\dagger {\lra{D}_\mu} \Sigma] \partial_\nu B^{\mu \nu}$   & $c_B$ & $\bm1$ & $c_B$ \\ \hline
  
    $ \frac{i }{ 2} c_W g\text{tr}[ \Sigma^\dagger \sigma_a {\lra{D}_\mu} \Sigma] D_\nu W_a^{\mu \nu}$   & $c_W$ & $\bm1$ & $c_W$ \\ \hline
    
        $ c_{y_u} \text{tr}[\Sigma^\dagger \Sigma] \sqrt{2} \bar{Q}_L \Sigma Y_u u_R$   & $c_{y_u}$ & $\bm1$ & $c_{y_u}$ \\ \hline
        
         $- i c_R^a \text{tr}[ \sigma^a  \Sigma^\dagger {\lra{D}_\mu} \Sigma] \bar{f}_R \gamma^\mu f_R$   & $ c_R^a $  & $\bm 3_R$ & $c_R \delta^{a,3}$ \\ \hline

         $- i c_L^a \text{tr}[ \sigma^a  \Sigma^\dagger {\lra{D}_\mu} \Sigma] \bar{f}_L \gamma^\mu f_L$   & $ c_L^a$  & $\bm 3_R$ & $c_L \delta^{a,3}$ \\ \hline

         $ i c_L^{(3)} \text{tr}[  \Sigma^\dagger  \sigma_a {\lra{D}_\mu} \Sigma] \bar{f}_L \sigma_a \gamma^\mu f_L$  & $c_L^{(3)}$ & $\bm1$ & $c_L^{(3)}$ \\  \hline
         
                  $- 4 i c_{R}^{ud}  \text{tr}[  \Sigma Y_d Y_u^{\dagger}  D_\mu \Sigma^\dagger] \bar{u}_R \gamma^\mu d_R$  & $c_{R}^{ud}$ & $\bm1$ & $c_{R}^{ud}$ \\

         \hline \hline
         
                  $ \kappa_{BB} g^{\prime a} g^{\prime a}  \text{tr}[\Sigma^\dagger \Sigma] B_{\mu \nu} B^{\mu \nu}$ & $\kappa_{BB}$  & $\bm1$ & $\kappa_{BB}$ \\ \hline
               $- i \kappa_{HB} g^{\prime a}  \text{tr}[\sigma^a D_\mu \Sigma^\dagger D_\nu \Sigma] B^{\mu \nu}$ & $\kappa_{HB}$  & $\bm1$ & $\kappa_{HB}$ \\ \hline
               
                $ i \kappa_{HW} g  \text{tr}[ D_\mu \Sigma^\dagger \sigma_a D_\nu \Sigma] W_a^{\mu \nu}$ & $\kappa_{HW}$ & $\bm1$ & $\kappa_{HW}$ \\ \hline    
                    $\kappa_{DB}  \sqrt{2}  \bar{Q}_L \Sigma Y_u \sigma^{\mu \nu} u_R B_{\mu \nu}$   & $\kappa_{DB}$ & $\bm1$ & $\kappa_{DB}$ \\ \hline    
                                                $\kappa_{DW}  \sqrt{2}  \bar{Q}_L \sigma^a \Sigma Y_u \sigma^{\mu \nu} u_R  W_{\mu \nu}^a$   & $\kappa_{DW}$ & $\bm1$ & $\kappa_{DW}$ \\ \hline    
                                                                                       $\kappa_{DG}  \sqrt{2}  \bar{Q}_L \Sigma Y_u  T^A \sigma^{\mu \nu} u_RG_{\mu \nu}^A $   & $\kappa_{DG}$ & $\bm1$ & $\kappa_{DG}$ \\ \hline                       
\end{tabular}
\caption{\it Transformation of the spurion Wilson coefficients of the $d=6$ operators under the custodial symmetry and their corresponding VEV. We are dropping fermion indices in the coefficients.}
\label{custoinv}
\end{table}
}


\end{document}